\pdfoutput=1

\documentclass[acmsmall,screen=true,natbib=true]{acmart_arxiv}
\usepackage{algorithm}
\usepackage{multirow}
\usepackage{cleveref}
\usepackage[noend]{algpseudocode}
\usepackage{subfigure}
\usepackage[export]{adjustbox}

\ifdefined\REVIEW
\setcopyright{acmcopyright}
\acmJournal{TODAES} 
\acmYear{0000} \acmVolume{x} \acmNumber{x} \acmArticle{x} \acmMonth{1} \acmPrice{15.00}\acmDOI{xx.xxxx/xxxxxxx}
\fi

\usepackage{tikz}

\ifdefined\REVIEW
\newcommand{\revise}[1]{\textcolor{blue}{#1}}
\else
\newcommand{\revise}[1]{\textcolor{black}{#1}}
\fi

\begin{document}
\ifdefined\REVIEW
\input{texts/cover-v1}
\clearpage
\setcounter{page}{1}
\fi
 
\title{Machine Learning for Electronic Design Automation: A Survey} 
\author{Guyue~Huang}
\authornote{These authors are ordered alphabetically.}
\author{Jingbo~Hu}
\authornotemark[1]
\author{Yifan~He}
\authornotemark[1]
\author{Jialong~Liu}
\authornotemark[1]
\author{Mingyuan~Ma}
\authornotemark[1]
\author{Zhaoyang Shen}
\authornotemark[1]
\author{Juejian~Wu}
\authornotemark[1]
\author{Yuanfan~Xu}
\authornotemark[1]
\author{Hengrui~Zhang}
\authornotemark[1]
\author{Kai~Zhong}
\authornotemark[1]
\author{Xuefei~Ning}
\email{nxf16@mails.tsinghua.edu.cn}
\affiliation{\institution{Tsinghua University} \country{China}}
\author{Yuzhe~Ma}
\author{Haoyu~Yang}
\author{Bei~Yu}
\email{byu@cse.cuhk.edu.hk}
\affiliation{\institution{Chinese University of Hong Kong} \country{Hong Kong SAR}}
\author{Huazhong~Yang}
\author{Yu~Wang}
\email{yu-wang@tsinghua.edu.cn}
\affiliation{\institution{Tsinghua University} \country{China}}



\renewcommand{\shortauthors}{G.~Huang et al.}
\begin{abstract}
With the down-scaling of CMOS technology, the design complexity of very large-scale integrated (VLSI) is increasing. Although the application of machine learning (ML) techniques in electronic design automation (EDA) can trace its history back to the 90s, the recent breakthrough of ML and the increasing complexity of EDA tasks have aroused more interests in incorporating ML to solve EDA tasks. In this paper, we present a comprehensive review of existing ML for EDA studies, organized following the EDA hierarchy.
\end{abstract}
 



\maketitle


\section{Introduction}
\label{sec:intro}

As one of the most important fields in applied computer/electronic engineering,
Electronic Design Automation (EDA) has a long history and is still under heavy development incorporating cutting-edge algorithms and technologies.
In recent years, with the development of semiconductor technology, the scale of integrated circuit (IC) has grown exponentially, challenging the scalability and reliability of the circuit design flow.
Therefore, EDA algorithms and software are required to be more effective and efficient to deal with extremely large search space with low latency.

Machine learning (ML) is taking an important role in our lives these days, which has been widely used in many scenarios. ML methods, including traditional and deep learning algorithms, achieve amazing performance in solving classification, detection, and design space exploration problems.
{Additionally, ML methods show great potential to generate high-quality solutions for many NP-complete (NPC) problems, which are common in the EDA field, while traditional methods lead to huge time and resource consumption to solve these problems. 
Traditional methods usually solve every problem from the beginning, with a lack of knowledge accumulation. 
Instead, ML algorithms focus on extracting high-level features or patterns that can be reused in other related or similar situations, avoiding repeated complicated analysis. 
Therefore, applying machine learning methods is a promising direction to accelerate the solving of EDA problems.}

In recent years, ML for EDA is becoming one of the trending topics, and a lot of studies that use ML to improve EDA methods have been proposed,
which cover almost all the stages in the chip design flow, {including design space reduction and exploration, logic synthesis, placement, routing, testing, verification, manufacturing, etc}. 
These ML-based methods have demonstrated impressive improvement compared with traditional methods.

We observe that most work collected in this survey can be grouped into four types:
\textit{decision making in traditional methods, performance prediction, black-box optimization, and automated design},
ordered by decreasing manual efforts and expert experiences in the design procedure, or an increasing degree of automation.
The opportunity of ML in EDA starts from \textit{decision making in traditional methods}, where an ML model is trained to select among available tool chains, algorithms,
or hyper-parameters, to replace empirical choice or brute-force search. 
ML is also used for \textit{performance prediction}, where a model is trained from a database of previously implemented designs to predict the quality of new designs,
helping engineers to evaluate new designs without the time-consuming synthesis procedure.
Even more automated, EDA tools utilized the workflow of \textit{black-box optimization},
where the entire procedure of design space exploration (DSE) is guided by a predictive ML model and a sampling strategy supported by ML theories.
Recent advances in Deep Learning (DL), especially Reinforcement Learning (RL) techniques have stimulated several studies that fully automate some complex design tasks with extremely large design space,
where predictors and policies are learned, performed, and adjusted in an online form,
showing a promising future of Artificial Intelligence (AI)-assisted \textit{automated design}.

This survey gives a comprehensive review of some recent important studies applying ML to solve some EDA important problems.
The review of these studies is organized according to their corresponding stages in the EDA flow.
Although the study on ML for EDA can trace back to the last century, most of the works included in this survey are in recent five years.
The rest of this survey is organized as follows.
In \Cref{sec:background}, we introduce the background of both EDA and ML.
From \Cref{sec:hls} to \Cref{sec:mask}, we introduce the studies that focus on different stages of the EDA flow,
i.e., high-level synthesis, logic synthesis \& physical design (placement and routing), and mask synthesis, respectively.
In \Cref{sec:analog}, analog design methods with ML are reviewed.
ML-powered testing and verification methods are discussed in \Cref{sec:test}.
Then, in \Cref{sec:misc}, other highly-related studies are discussed, including ML for SAT solver and the acceleration of EDA with deep learning engine.
The discussion of various studies from the ML perspective is given in \Cref{sec:ml}, which is complementary to the main organization of this paper.
Finally, \Cref{sec:conclu} concludes the existing ML methods for EDA and highlights future trends in this field.

\section{Background} \label{sec:background}\label{A2}
\subsection{Electronic Design Automation}

\begin{figure}[tb!]
    \centering
        \includegraphics[width=.44\linewidth]{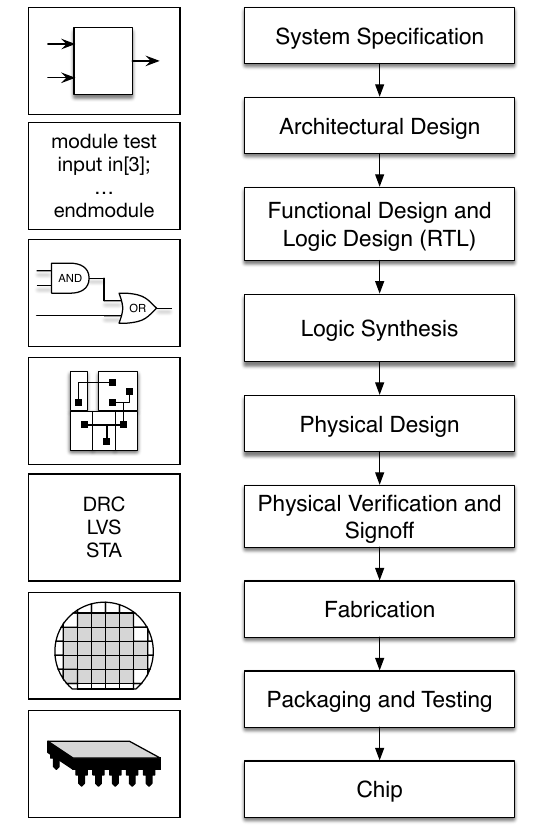}
    \caption{Modern chip design flow.}
    \label{fig:design-flow}
\end{figure}

Electronic design automation is one of the most important fields in electronic engineering.
In the past few decades, it has been witnessed that the flow of chip design became more and more standardized and complicated.
A modern chip design flow is shown in \Cref{fig:design-flow}.

High-level synthesis (HLS) provides automatic conversion from C/C++/SystemC-based specifications to hardware description languages (HDL). 
HLS makes hardware design much more convenient by allowing the designer to use high-level descriptions for a hardware system. 
However, when facing a large-scale system, HLS often takes a long time to finish the synthesis. 
Consequently, efficient design space exploration (DSE) strategy is crucial in HLS~\cite{Kim2018,Mahapatra2014,liu2013learning,Zuluaga2013,Meng2016}. 

Logic synthesis converts the behavioral level description to the gate level description, which is one of the most important problems in EDA.
Logic synthesis implements the specific logic functions by generating a combination of gates selected in a given cell library,
and optimizes the design for different optimization goals.
Logic synthesis is a complicated process that usually cannot be solved optimally, and hence the heuristic algorithms are widely used in this stage,
which include lots of ML methods~\cite{LSOracle,CNNlogic,RLGCNlogic,RLlogic}.

Based on the netlist obtained from synthesis, floorplanning and placement aim to assign the netlist components to specific locations on the chip layout. Better placement assignment implies the potential of better chip area utilization, timing performance, and routability.
Routing is one of the essential steps in very large-scale integrated (VLSI) physical design flow based on the placement assignment. Routing assigns the wires to connect the components on the chip. At the same time, routing needs to satisfy the requirements of timing performance and total wirelength without violating the design rules. The placement and routing are strongly coupled. Thus it is crucial to consider the routing performance even in the placement stage, and many ML-based routing-aware methods are proposed to improve the performance of physical design~\cite{pade_placer,Routenet,J-net,9045178,8394712,8533535}.


Fabrication is a complicated process containing multiple steps, which has a high cost in terms of time and resources. Mask synthesis is one of the main steps in the fabrication process, where lithography simulation is leveraged to reduce the probability of fabrication failure. Mask optimization and lithography simulation are still challenging problems. Recently, various ML-based methods are applied in the lithography simulation and mask synthesis~\cite{sarf-insertion,yang2018gan,mask-optimization,ye2019lithogan,DAMO}.

\revise{To ensure the correctness of a design, we need to perform design verification before manufacturing. In general, verification is conducted after each stage of the EDA flow, and the test set design is one of the major problems. Traditional random or automated test set generation methods are far away from optimal, therefore, there exist many studies that apply ML methods to optimize test set generation for verification~\cite{1219010, 4167997,4711612,1656859,7001424,5982005,Eder_ILP,6386595,4555820,Wang_GLSVISI18}.}

{After the chip design flow is finished, manufacturing testing needs to be carried out.} The chips need to go through various tests to verify their functionality and reliability.
{The coverage and the efficiency are two main optimization goals of the testing stage. Generally speaking, a large test set (i.e., a large number of test points) leads to higher coverage at the cost of high resource consumption.} To address the high cost of the testing process, studies have focused on applying ML techniques for test set optimization~\cite{4209884,5169847,8342115,9000131} and test {complexity} reduction~\cite{4079383,1443430,4358314}.


Thanks to decades of efforts from both academia and industry, the chip design flow is well-developed. However, with the huge increase {in} the scale of integrated circuits, more efficient and effective methods need to be incorporated to reduce the design cost.
Recent advancements in machine learning have provided a far-reaching data-driven perspective for problem-solving.
{In this survey, we review recent learning-based approaches for each stage in the EDA flow and also discuss the ML for EDA studies from the machine learning perspective.}

\subsection{Machine Learning} \label{bg:ml}

Machine learning is a class of algorithms that automatically extract information from datasets or prior knowledge.
Such a data-driven approach is a supplement to analytical models that are widely used in the EDA domain.
In general, ML-based solutions can be categorized according to their learning paradigms:
\textbf{supervised learning}, \textbf{unsupervised learning}, \textbf{active learning}, and \textbf{reinforcement learning}.
The difference between supervised and unsupervised learning is whether or not the input data is labeled.
With supervised or unsupervised learning, ML models are trained on static data sets offline and then deployed for online inputs without refinement.
With active learning, ML models subjectively choose samples from input space to obtain ground truth and refine themselves during the searching process.
With reinforcement learning, ML models interact with the environment by taking actions and getting rewards, with the goal of maximizing the total reward.
These paradigms all have been shown to be applied to the EDA problems.

As for the model construction, conventional machine learning models have been extensively studied for the EDA problems, especially for physical design~\cite{kahng2018new,zhuo2017accelerating}. 
Linear regression, random forest (RF)~\cite{liaw2002classification} and artificial neural networks (ANN)~\cite{hornik1989multilayer} are classical regression models.
Support vector machine (SVM)~\cite{boser1992training} is a powerful classification algorithm especially suitable for tasks with a small size of training set.
Other common classification models include K-Nearest-Neighbor (KNN) algorithm~\cite{fix1951discriminatory} and RF.
These models can be combined with ensemble or boosting techniques to build more expressive models.
For example, XGBoost~\cite{Chen_2016} is a gradient boosting framework frequently used in the EDA problems. 


Thanks to large public datasets, algorithm breakthrough, and improvements in computation platforms, there have been efforts of applying deep learning (DL) for EDA.
In particular, popular models in recent EDA studies include convolutional neural network (CNN)~\cite{ferianc2020improving,inference_sisa2018}, 
recurrent neural networks (RNN)~\cite{analog_edaa2016,learning}, generative adversarial network (GAN)~\cite{ye2019lithogan}, deep reinforcement learning (DRL)~\cite{mirhoseini2020chip,gcn-rl} and graph neural networks (GNN)~\cite{gcn-rl,zhang2019circuit}. CNN models are composed of convolutional layers and other basic blocks such as non-linear activation functions and down-sample pooling functions.
While CNN is suitable for feature extraction on grid structure data like 2-D image, RNN is good at processing sequential data such as text or audio. GNN is proposed for data organized as graphs. GAN trains jointly a generative network and a discriminative network which compete against each other to eventually generate high quality fake samples. DRL is a class of algorithms that incorporated deep learning into the reinforcement learning paradigm, where an agent learns a strategy from the rewards acquired with previous actions to determine the next action. DRL has achieved great success in complicated tasks with large decision space (e.g., Go game~\cite{silver2017mastering}).

\graphicspath{{image/hls/}}

\section{High level synthesis}
\label{sec:hls}

High-level synthesis (HLS) tools provide automatic conversion from C/C++/SystemC-based specification to hardware description languages like Verilog or VHDL.
HLS tools developed in industry and academia~\cite{10.1145/2514740,vivadohls,altera} have greatly improved productivity in customized hardware design.
High-quality HLS designs require appropriate pragmas in the high-level source code related to parallelism,
scheduling and resource usage, and careful choices of synthesis configurations in post-Register-Transfer-Level (RTL) stage.
Tuning these pragmas and configurations is a non-trivial task, and the long synthesis time for each design (hours from the source code to the final bitstream) prohibits exhaustive {DSE}. 

ML techniques have been applied to improve HLS tools from the following three aspects: fast and accurate result estimation
\cite{8457644,8892009,ferianc2020improving,yanghua2016improving,8714724, moham2019,Ustun2020accurate},
refining conventional DSE algorithms~\cite{Kim2018,Mahapatra2014, Wang2020machine},
and reforming DSE as an active-learning problem~\cite{7577370,liu2013learning,Zuluaga2013,Meng2016}. 
In addition to achieving good results on individual problems, previous studies have also introduced new generalizable techniques about feature engineering
\cite{8457644,8892009,yanghua2016improving,8714724, moham2019},
selection and customization of ML models~\cite{Ustun2020accurate},
and design space sampling and searching strategies~\cite{liu2013learning,Zuluaga2013,Meng2016}.


This section is organized as follows. 
\Cref{sec:hls:prediction} introduces recent studies on employing ML for result estimation, often in a static way.
\Cref{sec:hls:dse} introduces recent studies on adopting ML in DSE workflow, either to improve conventional methods or in the form of active learning. 

\subsection{Machine Learning for Result Estimation}
\label{sec:hls:prediction}
{The reports from HLS tools provide important guidance for tuning the high-level directives. However, acquiring accurate result estimation in an early stage is difficult due to complex optimizations in the physical synthesis, imposing a trade-off between accuracy (waiting for post-synthesis results) and efficiency (evaluating in the HLS stage).} 
ML can be used to improve the accuracy of HLS reports through learning from real design benchmarks.
In \Cref{sec:hls:prediction:report}, we introduce previous work on predicting the timing, resource usage, {and operation delay} of an HLS design. In \Cref{sec:hls:prediction:crossplatform} we describe two types of research about cross-platform performance prediction.

\subsubsection{Estimation of Timing, Resource Usage, and Operation Delay} \label{sec:hls:prediction:report}

\begin{figure}[tb!]
    \centering
    \includegraphics[width=.426\textwidth]{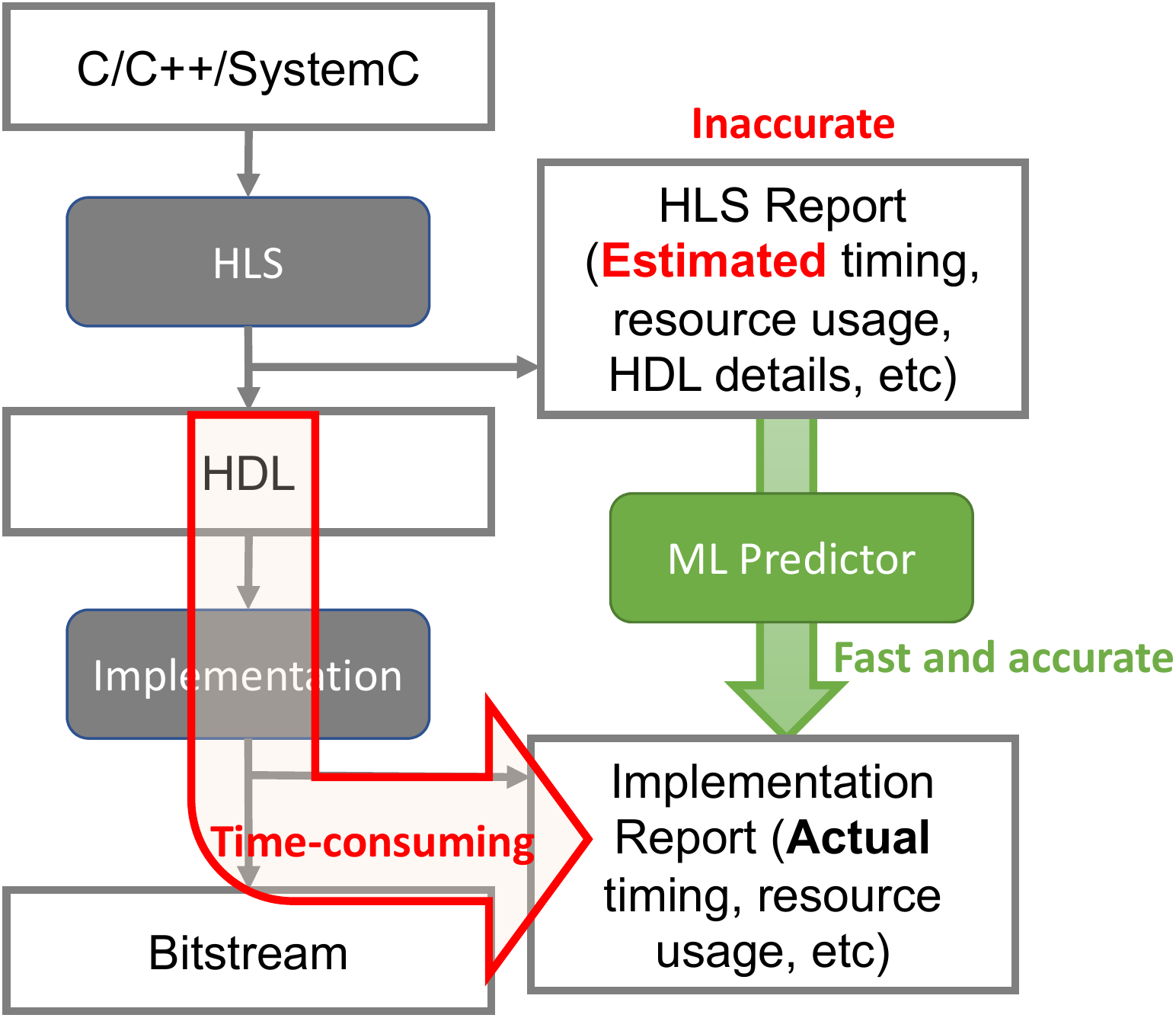}
    \caption{FPGA tool flow with HLS, highlighting ML-based result predictor (reproduced from \cite{8457644}).}
    \label{fig:prediction_workflow}
\end{figure}

The overall workflow of timing and resource usage prediction is concluded in \Cref{fig:prediction_workflow}.
This workflow is first proposed by \citet{8457644} and augmented by \citet{8892009} and \citet{ferianc2020improving}. The main {methodology} is to train an ML model that takes HLS reports as input and outputs a more accurate implementation report without conducting the time-consuming post-implementation. The workflow proposed by \citet{8457644} can be divided into two steps: \emph{data processing} and \emph{training estimation models}.

\textbf{Step 1: Data Processing}. To enable ML for HLS estimation, we need a dataset for training and testing. The HLS and implementation reports are usually collected across individual designs by running each design through the complete C-to-bitstream flow, for various clock periods and targeting different FPGA devices. After that, {one can} extract features from the HLS reports as inputs and features from implementation reports as outputs. Besides, to overcome the effect of colinearity and reduce the dimension of the data, {previous studies} often apply feature selection techniques to systematically remove unimportant features.
The most commonly used features are summarized in \Cref{hls:table1}.
\begin{table}[htb]
\centering
\caption[tablehls1]{Categories of selected features and descriptions~\cite{8457644,  8892009}}
\label{hls:table1}
\footnotesize
\begin{tabular}{p{2.5cm}p{7.6cm}}
  \toprule
      Category & Brief Description \\
  \midrule
      Clock periods & Target clock period; achieved clock period \& its uncertainty. \\
      Resources & Utilization and availability of LUT, FF, DSP, and BRAM. \\
      Logic Ops & Bitwidth/resource statistics of operations. \\
      Arithmetic Ops & Bitwidth/resource statistics of arithmetic operations. \\
      Memory & Number of memory words/banks/bits; resource usage for memory. \\
      Multiplexer & Resource usage for multiplexers; multiplexer input size/bitwidth. \\
  \bottomrule
\end{tabular}
\end{table}

\textbf{Step 2: Training Estimation Models}. After constructing the dataset, {regression models are trained to estimate post-implementation resource usages and clock periods.}
Frequently used metrics to report the estimation error include relative absolute error (RAE) and relative root mean squared error (RMSE).
For both metrics, lower is better.
RAE is defined in \Cref{RAE}, where $\hat{y}$ is a vector of values predicted by the model, $y$ is a vector of actual ground truth values in the testing set, and $\bar{y}$ denotes the mean value of $y$.
\begin{equation}\label{RAE}
    \text{RAE} = \frac{| \hat{y} - y |}{| y - \overline{y} |}.
  \end{equation}
  
Relative RMSE is given by \Cref{RMSE}, where $N$ is the number of samples, and $\hat{y}_i$ and $y_i$ are the predicted and actual values of a sample, respectively.
\begin{equation}\label{RMSE}
    \text{Relative RMSE} = \sqrt{\frac{1}{N}\sum^N_{i=1} (\frac{\hat{y}_i-y_i}{y_i})^2}\times {100\%}.
\end{equation}


\citet{8892009} model timing as a regression problem, and use the Minerva tool~\cite{8279804} to obtain results in terms of maximum clock frequency, throughput, and throughput-to-area ratio for the RTL code generated by the HLS tool.
Then an ensemble model combining linear regression, neural network, SVM, and random forest,
is proposed to conduct estimation and achieve an accuracy higher than 95\%.
There are also studies that predict {whether} a post-implementation is required or not, instead of predicting the implementation results.
{As} a representative study, \citet{7577370} train a predictive model to avoid re-synthesizing each new configuration.

{ML techniques have been applied recently to reduce the HLS tool's prediction error of the operation delay~\cite{Ustun2020accurate}. Existing HLS tools perform delay estimations based on the simple addition of pre-characterized delays of individual operations, and can be inaccurate because of the post-implementation optimizations (e.g., mapping to hardened blocks like DSP adder cluster). A customized Graph Neural Network (GNN) model is built to capture the association between operations from the dataflow graph, and train this model to infer the mapping choices about hardened blocks. Their method can reduce the RMSE of the operation delay prediction of Vivado HLS by 72\%.}

\subsubsection{Cross-Platform Performance Prediction} \label{sec:hls:prediction:crossplatform}

\begin{figure}[htb]
    \centering
    \includegraphics[width=.78\linewidth]{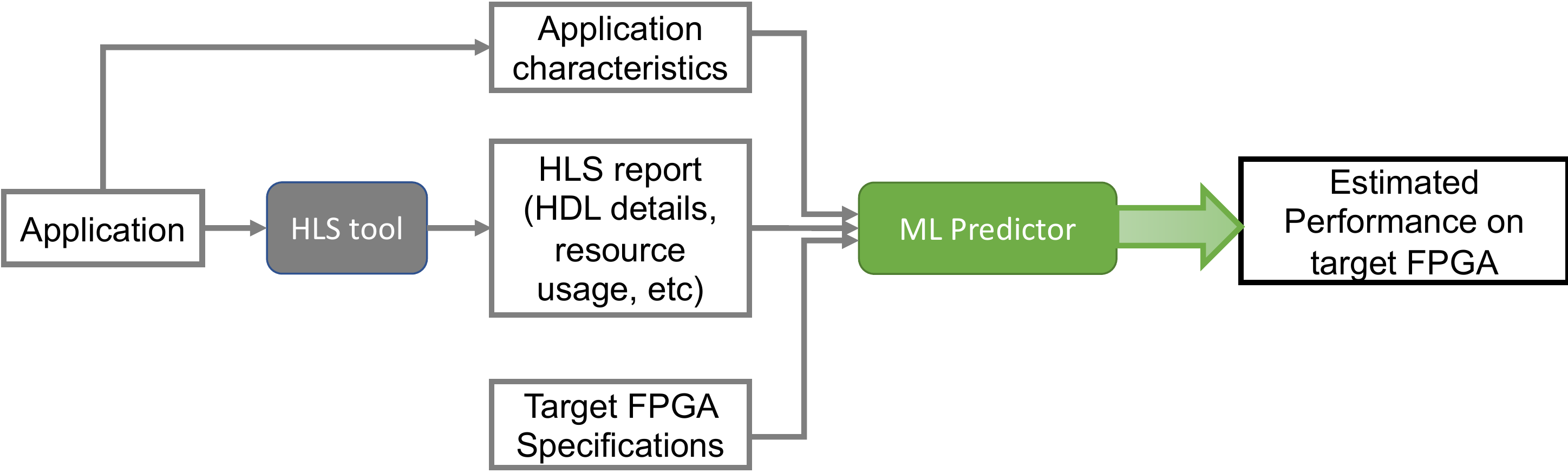}
    \caption{Overall workflow of XPPE (reproduced from \cite{moham2019}).}
    \label{fig:XPPE}
\end{figure}

Hardware/software co-design enables designers to take advantage of new hybrid platforms such as Zynq.
However, dividing an application into two parts {makes} the platform selection difficult for the developers, {since} there is a huge variation in the application’s performance of the same workload across various platforms. To avoid fully implementing the design on each platform, \citet{moham2019} propose an ML-based cross-platform performance estimator, XPPE, and its overall workflow is described in \Cref{fig:XPPE}.
The key {functionality} of XPPE is using the resource utilization of an application on one specific FPGA to estimate its performance on other FPGAs. 

XPPE uses a Neural Network (NN) {model} to estimate the speedup of an application for a target FPGA over the ARM processor.
The inputs of XPPE are available resources on target FPGA, resource utilization report {from} HLS Vivado tool (extracted features, similar to the features in \Cref{hls:table1}), and application's characteristics. The output is the speedup estimation on the target FPGA over an ARM A-9 processor.
This method is similar {to} \citet{8457644} and \citet{8892009} in that they all take the features in HLS reports as input and {aim to avoid} the time-consuming post-implementation.
The main difference is that the input and output {features} in XPPE are from different platforms.
{The relative RMSE between the predictions and the real measurements is used to evaluate the accuracy of the estimator}.
The proposed architecture can achieve a relative mean square error of 5.1\% and the speedup is more than 0.98$\times$.

Like XPPE, \citet{8587690} also propose an ML-based cross-platform estimator, named HLSPredict. There are two differences. First, HLSPredict only takes workloads (the applications in XPPE) as inputs instead of the combination of HLS reports, application's characteristics and specification of target FPGA devices. Second, the target platform of HLSPredict must be the same as the platform in the training stage. In general, HLSPredict aims to rapidly estimate performance on a specific FPGA by direct execution of a workload on a commercially available off-the-shelf host CPU, but XPPE aims to accurately predict the speedup of different target platforms. 
For optimized workloads, HLSPredict achieves a relative absolute percentage error ($APE=|\frac{y-\hat{y}}{y}|$) of 9.08\% and a 43.78$\times$ runtime speedup compared with FPGA synthesis and direct execution.




\subsection{Machine Learning for Design Space Exploration in HLS}
\label{sec:hls:dse}

In the previous subsection, we describe how ML models are used to predict the quality of results.
Another application of ML in HLS is to assist DSE.
The tunable synthesis options in HLS, provided in the form of pragmas, span a very large design space.
Most often, the task of DSE is to find the Pareto Frontier Curve, {on which every point is not fully dominated by any other points under all the metrics}.


{Classical search algorithms have been applied in HLS DSE, such as Simulated Annealing (SA) and Genetic Algorithm (GA). But these algorithms are unable to learn from the database of previously explored designs. Many previous studies use an ML predictive model to guide the DSE.
The models are trained on the synthesis results of explored design points, and used to predict the quality of new designs (See more discussions on this active learning workflow in \Cref{sec:ml:models}).
Typical studies are elaborated in \Cref{sec:hls:dse:learning}.
There is also a thread of work that involves learning-based methods to improve the inefficient or sensitive part of the classical search algorithms, as elaborated in \Cref{sec:hls:dse:traditional}.
Some work included in this subsection focuses on system-level DSE rather than HLS design~\cite{Kim2018}, or general active learning theories~\cite{Zuluaga2013}.}



\begin{figure}
    \centering
    \includegraphics[width=250pt]{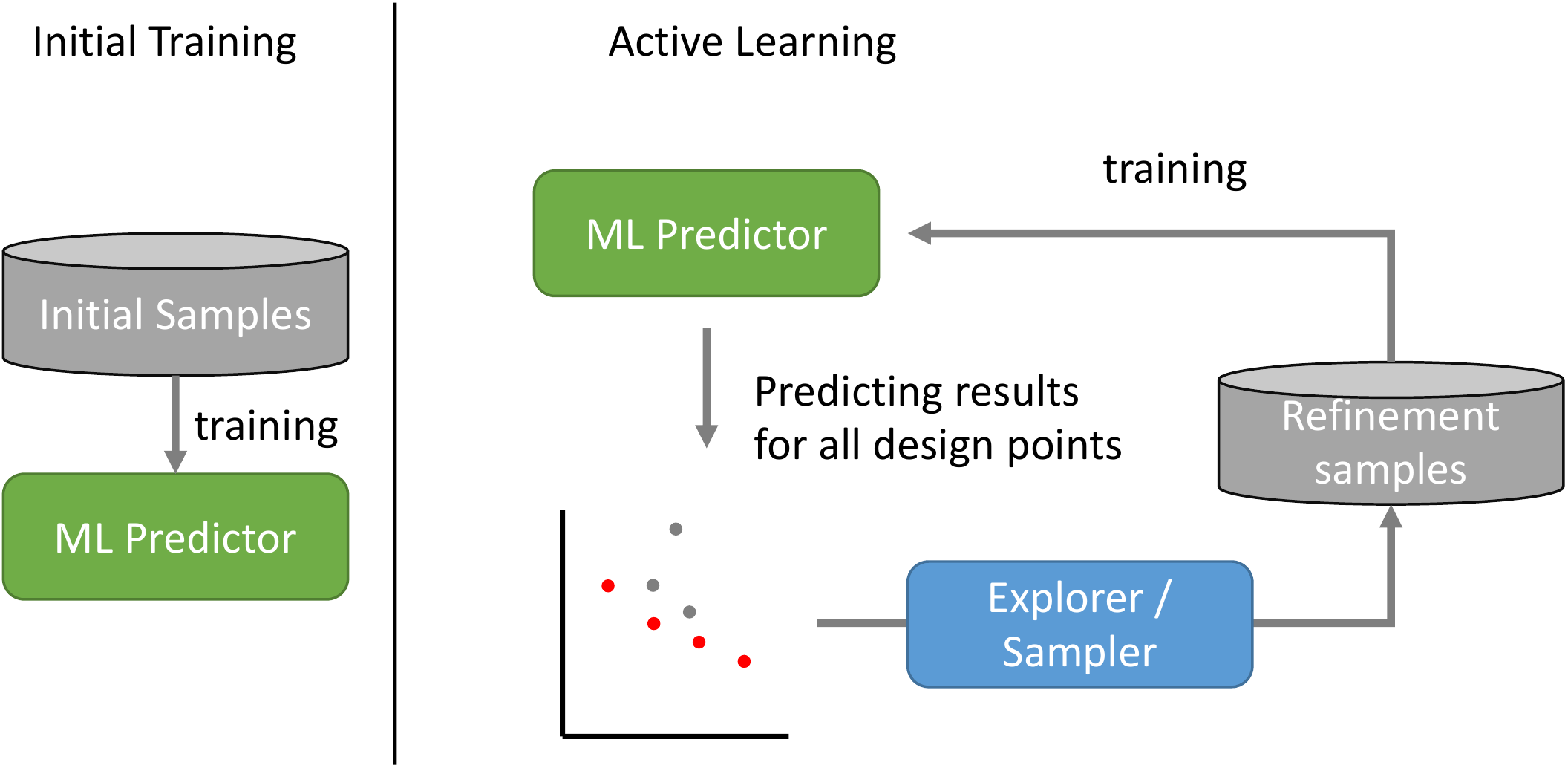}
    \caption{The iterative-refinement DSE framework (reproduced from ~\cite{liu2013learning}).}
    \label{fig:dseframe}
\end{figure}


\subsubsection{{Active Learning}} \label{sec:hls:dse:learning}
The four papers visited in this part utilize the active learning approach to perform DSE for HLS,
and use predictive ML models to surrogate actual synthesis when evaluating a design.
\citet{7577370} propose a design space explorer that selects new designs to implement through an active learning approach.
Transductive experimental design {(TED)}~\cite{liu2013learning} focuses on seeking the samples that describe the design space accurately.
Pareto active learning {(PAL)} in~\cite{Zuluaga2013} is proposed to sample designs which the learner cannot clearly {classify}.
Instead of focusing on how accurately the model describes the design space,
adaptive threshold non-pareto elimination {(ATNE)}~\cite{Meng2016} estimates the inaccuracy of the learner and achieves better performance than TED and PAL. 

\citet{7577370} propose a dedicated explorer to search for Pareto-optimal HLS designs for FPGAs.
The explorer iteratively selects potential Pareto-optimal designs to synthesize and verify.
The selection is based on a set of important features, which are adjusted during the exploration.
The proposed method runs 6.5$\times$ faster than an exhaustive search, and runs 3.0$\times$ faster than a restricted search method but finds results with higher quality.

{The basic idea of TED~\cite{liu2013learning} is} to select representative as well as the hard-to-predict samples from the design space, instead of the random sample used in previous work.
{The target is to maximize the accuracy of the predictive model with the fewest training samples. The authors formulate the problem of finding the best sampling strategy as follows:}  
TED assumes that the overall number of knob settings is $n$($|\mathcal{K}|=n$), from which we want to select a training set ${\tilde{\mathcal{K}}}$ such that $|{\tilde{\mathcal{K}}}|=m$. Minimizing the prediction error $H(k)-\tilde{H}(k)$ for all $k\in \mathcal{K}$ is equivalent to the following problem:
$$
\max_{{\tilde{\mathcal{K}}}}T[\mathcal{K}{\tilde{\mathcal{K}^T}}({\tilde{\mathcal{K}}}{\tilde{\mathcal{K}^T}}+\mu I)^{-1}{\tilde{\mathcal{K}}}\mathcal{K^T}]\ s.t.\ {\tilde{\mathcal{K}}}\subset\mathcal{{K}},|{\tilde{\mathcal{K}}}|=m,
$$
where $T[\cdot]$ is the matrix trace operator and $\mu>0$.
The authors interpret their solution as sampling from a set ${\tilde{\mathcal{K}}}$ that \textit{span a linear space, to retain most of the information of $\mathcal{K}$}~\cite{liu2013learning}.

PAL~\cite{Zuluaga2013} {is proposed for general active learning scenarios and is demonstrated by a sorting network synthesis DSE problem in the paper. It} uses Gaussian Process (GP) to predict Pareto-optimal points in design space . The models predict the objective functions to identify points that are Pareto-optimal with high probabilities. A point $x$  that has not been sampled is predicted as $\hat{f}(x)=\mu(x)$ and $\sigma(x)$ is interpreted as the uncertainty of the prediction which can be captured by the hyperrectangle
$$
Q_{\mu, \sigma, \beta}(x)=\left\{y: \mu(x)-\beta^{1 / 2} \sigma(x) \preceq y \preceq \mu(x)+\beta^{1 / 2} \sigma(x)\right\},
$$
where $\beta$ is a scaling parameter to be chosen.
{PAL focuses on accurately predicting points near the Pareto frontier, instead of the whole design space.
In every iteration, the algorithm classifies samples into three groups: Pareto-optimal, Non-Pareto-optimal, and uncertain ones.}
The next design point to evaluate is the one with the largest uncertainty, which intuitively has more information to improve the model.
The training process is terminated when { there are no uncertain points}.
The points classified as Pareto-optimal are then returned.


{ATNE~\cite{Meng2016} utilizes Random Forest (RF) to aid the DSE process. This work uses a Pareto identification threshold that adapts to the estimated inaccuracy of the RF regressor and eliminates the non-Pareto-optimal designs incrementally.}
Instead of focusing on improving the accuracy of the learner, ATNE focuses on {estimating and minimizing the risk of losing ``good''} designs due to learning inaccuracy.

\subsubsection{{Machine Learning for Improving Other Optimization Algorithms}} \label{sec:hls:dse:traditional}
In this part, we {summarize three studies that use ML techniques to improve classical optimization algorithms}.


{STAGE~\cite{Kim2018} is proposed for DSE of many-core systems. The motivating observation of STAGE is that the performance of simulated annealing is highly sensitive to the starting point of the search process.} The authors build an ML model to learn which parts of the design space should be focused on, eliminating the times of {futile} exploration~\cite{10.1162/15324430152733124}.
{The proposed strategy is divided into two stages. }The first stage (local search) performs a normal local search, guided by a cost function based on the designer’s goals. The second stage (meta search) tries to use the search trajectories from previous local search runs to learn to predict the outcome of local search given a certain starting point~\cite{Kim2018}.

\begin{figure}
    \centering
    \includegraphics[width=.48\linewidth]{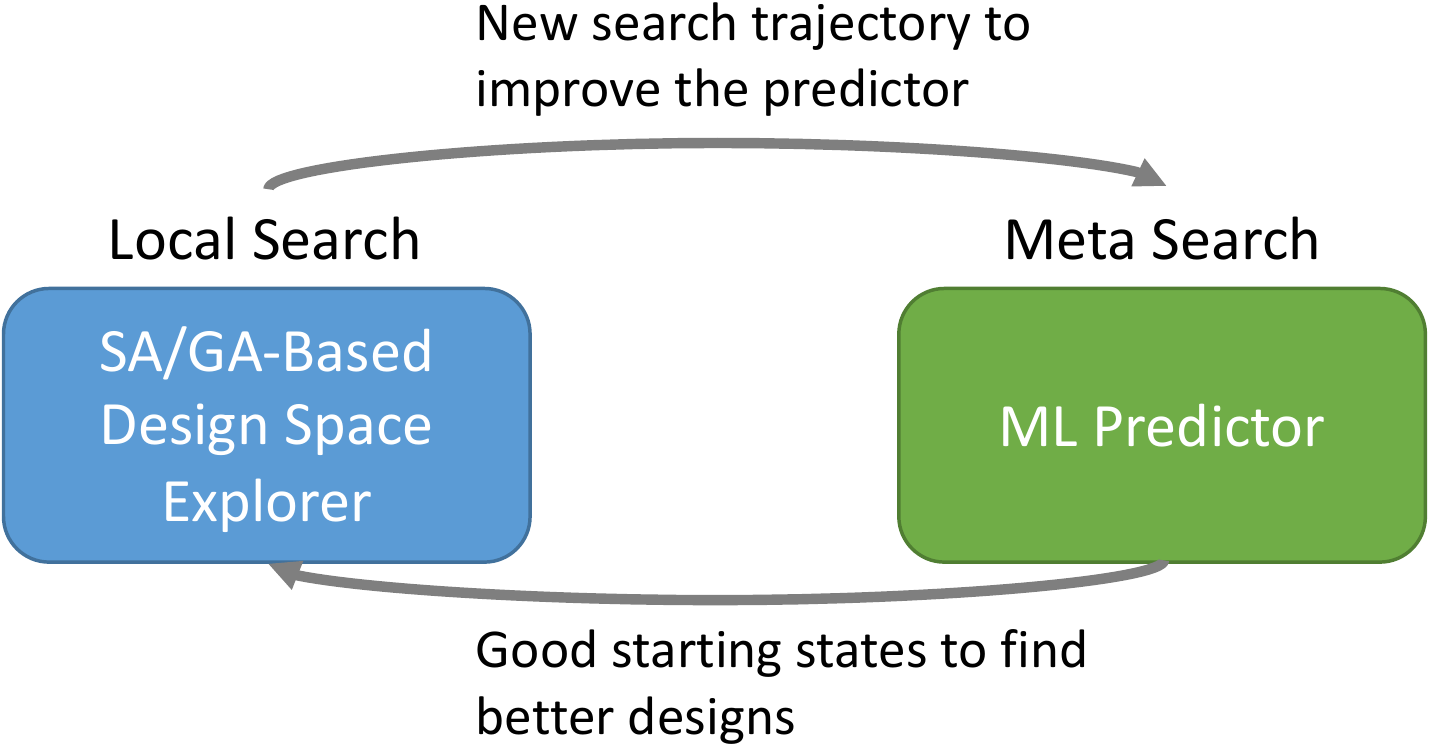}
    \caption{Overview of the STAGE algorithm (reproduced from \cite{Kim2018}).}
    \label{fig:stage}
\end{figure}

Fast Simulated Annealing {(FSA)}~\cite{Mahapatra2014} utilizes the decision tree to improve the performance of {SA}. Decision tree learning is a widely used method for inductive inference. {The HLS pragmas are taken as input features.} FSA {first performs} standard SA to generate enough training sets to build the decision tree. Then it generates new design configurations with the decision tree and keeps the dominating designs~\cite{Mahapatra2014}.

{In a recent study, \citet{Wang2020machine} propose several ML techniques to help decide the hyper-parameter settings of three meta-heuristic algorithms: SA, GA and Ant Colony Optimizations (ACO). For each algorithm, the authors build an ML model that predicts the resultant design quality (measured by Average Distance to the Reference Set, ADRS) and runtime from hyper-parameter settings. Compared with the default hyper-parameters, their models can improve the ADRS by more than 1.92$\times$ within similar runtime. The authors also combine SA, GA and ACO to build a new design space explorer, which further improves the search efficiency.}

\begin{table*} 
    \centering
    \caption{{Summary of ML for HLS} }
    \label{tb:hls}
    \footnotesize
    \begin{tabular}{p{3cm}|p{4cm}|p{3cm}|p{1cm}}
            \toprule 
            \centering{\textbf{Task}} & \textbf{Task Details} & \textbf{ML Algorithm} & \textbf{Reference} \\ 
            \midrule 
            \multirow{4}{3cm}{\centering Result prediction} 
            & Timing and resource usage prediction & Lasso, ANN, XGBoost & \cite{8457644} \\ 
            \cmidrule(lr){2-4}
            & Max frequency, throughput, area & Ridge regression, ANN, SVM, Random Forest & \cite{8892009} \\
            \cmidrule(lr){2-4}
            & Latency & Gaussian Process & \cite{ferianc2020improving} \\
            \cmidrule(lr){2-4}
            & Operation delay & Graph Neural Network & \cite{Ustun2020accurate} \\
            \cmidrule(lr){1-4}
            \multirow{2}{3cm}{\centering{Cross-platform}} 
            & Predict for new FPGA platforms & ANN & \cite{moham2019}\\
            \cmidrule(lr){2-4}
            & Predict for new applications through executing on CPUs & Linear models, ANN, Random Forest & \cite{8587690} \\
            \cmidrule(lr){1-4}
            \multirow{3}{3cm}{\centering{Active learning }}
            & Reduce prediction error with fewer samples & Random Forest, Gaussian Process Regression & \cite{liu2013learning} \\
            \cmidrule(lr){2-4}
            & Reduce prediction error for points near the Pareto-frontier & Gaussian Process & \cite{Zuluaga2013} \\
            \cmidrule(lr){2-4}
            & Reduce the risk of losing Pareto designs  & Random Forest & \cite{Meng2016} \\
            \cmidrule(lr){1-4}
            \multirow{3}{3cm}{\centering{Improving conventional algorithms}} 
            & Initial point selection & Quadratic regression  & \cite{Kim2018} \\
            \cmidrule(lr){2-4}
            & Generation of new sample & Decision Tree & \cite{Mahapatra2014} \\
            \cmidrule(lr){2-4}
            & Hyper-parameter selection & Decision Tree & \cite{Wang2020machine} \\
            \bottomrule
            
        \end{tabular} 
\end{table*}

\subsection{Summary of Machine Learning for HLS}
{This section reviews recent work on ML techniques in HLS, as listed in \Cref{tb:hls}. Using ML-based timing/resource/latency predictors and data-driven searching strategies, the engineering productivity of HLS tools can be further improved and higher-quality designs can be generated by efficiently exploring a large design space.}

We believe the following practice can help promote future research of ML in HLS:
\begin{itemize}
    \item Public benchmark for DSE problems. The researches about result estimation are all evaluated on public benchmarks of HLS applications, such as Rosetta~\cite{10.1145/3174243.3174255}, MachSuite~\cite{6983050}, etc. However, DSE researches are often evaluated on a few applications because the cost of synthesizing a large design space for each application is heavy. Building a benchmark that collects different implementations of each application can help fairly evaluate DSE algorithms.
    \item Customized ML models. Most of the previous studies use off-the-shelf ML models. Combining universal ML algorithms with domain knowledge can potentially improve the performance of the model. For example, \citet{Ustun2020accurate} customize a standard GNN model to handle the specific delay prediction problem, which brings extra benefit in model accuracy.
\end{itemize}

\section{Logic Synthesis and Physical design}
\label{sec:physical}




In the logic synthesis and physical design stage, there are many key sub-problems that can {benefit from} the power of ML models, {including} lithography hotspot detection, path classification, congestion prediction, placement guide, fast timing analysis, logic synthesis scheduling, and so on.
{In this section, we organize the review of studies by their targeting problems.}



\subsection{Logic Synthesis} \label{sec:physical:logicsynthesis}

Logic synthesis is an optimization problem with complicated constraints, which requires accurate solutions. Consequently, using ML algorithms to directly generate logic synthesis solutions is difficult. However, there are some {studies} using ML algorithms to schedule existing traditional optimization strategies.
For logic synthesis, LSOracle~\cite{LSOracle} relies on DNN to dynamically decide which optimizer should be applied to different {parts} of the circuit.
The framework exploits two optimizers, and-inverter graph (AIG) and majority-inverter graph (MIG),
and applies k-way partitioning on circuit {directed acyclic graph (DAG)}.

There are many logic transformations in current synthesis tools such as ABC~\cite{abc}.
To select an appropriate synthesis flow, \citet{CNNlogic} formulate a multi-class classification problem and design a CNN to map a synthesis flow to quality of results (QoR) levels.
The prediction on unlabeled flows are then used to select the optimal synthesis flow.
The CNN takes the one-hot encoding of synthesis flows as inputs and outputs the possibilities of the input flow belonging to different QoR metric levels.

Reinforcement learning is also employed for logic synthesis in \cite{RLGCNlogic,RLlogic}.
A transformation between two DAGs with the same I/O behaviors is modeled as an action.
In \cite{RLGCNlogic}, GCN is utilized as a policy function to obtain the probabilities for every {action}.
\cite{RLlogic} employs advantage actor critic agent (A2C) to search the optimal solution.

\subsection{Placement and Routing Prediction} \label{sec:physical:PnR}

\subsubsection{Traditional Placers Enhancement} \label{sec:physical:PnR:traditional}
While previous fast placers can conduct random logic placement efficiently with good performances, researchers find that their placement of data path logic is suboptimal. PADE~\cite{pade_placer} proposes a placement process with automatic data path extraction and evaluation, in which the placement of data path logic is conducted separately from random logic. PADE is a force-directed global placer, which applies SVM and NN to extract and evaluate the data path patterns with high dimensional data such as netlist symmetrical structures, initial placement hints, and relative area. The extracted data path is mapped to bit stack structure and uses SAPT~\cite{keep-it-straight} (a placer placed on SimPL~\cite{SimPL}) to optimize separately from random logic.


\subsubsection{Routing Information Prediction} \label{sec:physical:PnR:routingprediction}

The basic requirements of routing design rules must be considered in the placement stage.
However, it is difficult to predict routing information in the placement stage accurately and fast, and researchers recently employ machine learning to solve this.
RouteNet~\cite{Routenet} is the first work to employ CNN for design rule checking (DRC) hotspot detection.
The input features of a customized fully convolutional network (FCN) include the outputs of rectangular uniform wire density (RUDY), a pre-routing congestion estimator.
An 18-layer ResNet is also employed to predict design rule violation (DRV) count.
A recent work~\cite{J-net} abstracts the pins and macros density in placement results into image data,
and utilizes a pixel-wise loss function to optimize an encoder-decoder model (an extension of U-Net architecture).
The network output is a heat-map, which represents the location where detailed routing congestion may occur.
PROS \cite{ROUTE-ICCAD2020-Chen} takes advantages of fully convolution networks to predict routing congestion from global placement results.
The framework is demonstrated efficient on industrial netlists.
\citet{pui2017clock} explore the possibilities using ML methods to predict routing congestion in UltraScale FPGAs.
{\citet{9045178} transfer} the routing congestion problem in large-scale FPGAs to an image-to-image problem, and then uses conditional GAN to solve it.
In addition, there are some studies that only predict the number of congestions instead of the location of congestion~\cite{8394712,8533535}.
\citet{8533535} use models like linear regression, RF and MLP to learn how to use features from earlier stages to produce more accurate congestion prediction, so that the placement strategy can be adjusted.
\citet{ML4DR} predict the detailed routing congestion using nonparametric regression algorithm,
multivariate adaptive regression splines (MARS) with the global information as inputs.
Another study~\cite{ICCD2016} takes the netlist, clock period, utilization, aspect ratio and BEOL stack as inputs and utilizes MARS and SVM to predict the routability of a placement.
This study also predicts Pareto frontiers of utilization, number of metal layers, and aspect ratio.
Study in \cite{PLACE-DATE2021-Liu} demonstrates the potential of embedding ML-based routing congestion estimator into global placement stage.
Recently, \citet{crosstalk} build a routing-free crosstalk prediction model by adopting several ML algorithms such as regression, NN, GraphSAGE and GraphAttention.
The proposed framework can identify nets with large crosstalk noise before the routing step, which allows us to modify the placement results to reduce crosstalk in advance.

There is also a need to estimate the final wirelength, timing performance, circuit area, power consumption, clock and other parameters in the early stage.
Such prediction task can be modeled as a regression task and commonly-used ML models include SVM, Boosting, RF, MARS, etc.
\citet{Jeong} learn a model with MARS to predict performance from a given set of circuit configurations, with NoC router,
a specific functional circuit and a specific business tool.
In \cite{8715016}, the researchers introduce linear discriminant analysis (LDA) algorithm to find seven combined features for the best representation,
and then a KNN-like approach is adopted to combine the prediction results of ANN, SVM, LASSO, and other machine learning models.
In this way, {\citet{8715016}} improve the wirelength prediction given by the virtual placement and routing in the synthesis.
{\citet{8394712}} predict the final circuit performance in the macro placement stage, and {\citet{7835438}} predict the circuit performance {in the global routing stage}, including congestion number, hold slack, area and power. 

For sign-off timing analysis, {\citet{8807063} use} random forest to give the sign-off timing slack from hand-crafted features.
{Another research}~\cite{6681682} works on sign-off timing analysis and use linear regression to fit the static timing analysis (STA) model,
thus reduce the frequency that the incremental static timing analysis (iSTA) tool need to be called.
{\citet{6800474} propose SI for Free, a regression method to predict expensive signal integrity (SI) mode sign-off timing results by using cheap non-SI mode sign-off timing analysis.
{\cite{7171706} propose} golden timer extension (GTX), a framework to reduce mismatches between different sign-off timing analysis tools to obtain neither optimistic nor pessimistic results.}

{\citet{GAN-CTS} employ GAN and RL for clock tree prediction.
Flip flop distribution, clock net distribution, and trial routing results serve as input images.
For feature extraction, GAN-CTS adopts transfer learning from a pre-trained ResNet-50 on the ImageNet dataset by adding fully-connected (FC) layers.
A conditional GAN is utilized to optimize the clock tree synthesis, of which the generator is supervised by the regression model.
An RL-based policy gradient algorithm is leveraged for the clock tree synthesis optimization.}

\subsubsection{Placement Decision Making} \label{sec:physical:PnR:placementdecision}

As the preliminary step of the placement, floorplanning aims to roughly determine the geometric relationship among circuit modules and to estimate the cost of the design.
He et al.~\cite{he2020learn} explore the possibility of acquiring local search heuristics through a learning mechanism.
More specifically, an agent has been trained using a novel deep Q-learning algorithm to perform a walk in the search space by selecting a candidate neighbor solution at each step, while avoiding introducing too much prior human knowledge during the search.
Google~\cite{mirhoseini2020chip} recently models chip placement as a sequential decision making problem and trains an RL policy to make placement decisions.
During each episode, the RL agent lays the macro in order.
After arranging macros, it utilizes the force-directed method for standard cell placement.
GCN is adopted in this work to embed information related to macro features and the adjacency matrix of the netlist.
Besides, FC layers are used to embed metadata.
After the embedding of the macros, the graph and the metadata, another FC layer is applied for reward prediction.
Such embedding is also fed into a deconvolution CNN model, called PolicyNet, to output the mask representing the current macro placement.
The policy is optimized with RL to maximize the reward, which is the weighted average of wirelength and congestion.

\subsection{Power Deliver Network Synthesis and IR Drop Predictions}

Power delivery network (PDN) design is a complex iterative optimization task, which strongly influences the performance, area and cost of a chip.
To reduce the design time, recent studies have paid attention to ML-based IR drop estimation, a time-consuming sub-task.
Previous work usually adopts simulator-based IR analysis, which is challenged by the increasing complexity of chip design.
IR drop can be divided into two categories: static and dynamic.
Static IR drop is mainly caused by voltage deviation of the metal wires in the power grid,
while dynamic IR drop is led by the switching behaviors and localized fluctuating currents.
In IncPIRD~\cite{IncPIRD}, the authors employ XGBoost to conduct incremental prediction of static IR drop problem,
which is to predict IR value changes caused by the modification of the floorplan.
For dynamic IR drop estimation, \citet{PowerNet} aim to predict the IR values of different locations and models IR drop estimation problem as a regression task.
This work introduces a ``maximum CNN'' algorithm to solve the problem.
Besides, PowerNet is designed to be transferable to new designs, while most previous studies train models for specific designs.
A recent work~\cite{GridNet} proposes an electromigration-induced IR drop analysis framework based on conditional GAN.
The framework regards the time and selected electrical features as input images and outputs the voltage map.
Another recent work~\cite{Template} focuses on PDN synthesis in floorplan and placement stages.
This paper designs a library of stitchable templates to represent the power grid in different layers.
In the training phase, SA is adopted to choose a template.
In the inference phase, MLP and CNN are used to choose the template for floorplan and placement stages, respectively.
\citet{10.1145/3287624.3287689} use hybrid surrogate modeling (HSM) that combines SVM, ANN and MARS to predict the bump inductance that represents the quality of the power delivery network.

\begin{table}
    \centering
    \caption{{Summary of ML for logic synthesis and physical design}}
    \label{tb::physical_summary}
    	\footnotesize
\begin{tabular}{p{2.5cm}|p{7cm}|p{2cm}|p{0.6cm}}
\toprule
     \textbf{\centering{Section}} & \textbf{Task} & \textbf{ML Algorithm} & \textbf{Reference}\\
\midrule
    \multirow{4}{2.5cm}{\centering Logic Synthesis}
    & To decide which optimizer (AIG/MIG) should be utilized for different circuits.  & DNN  & \cite{LSOracle} \\ \cmidrule(lr){2-4}
    & To classify the optimal synthesis flows. & CNN     & \cite{CNNlogic}  \\ \cmidrule(lr){2-4}
    & To generate the optimal synthesis flows. & GCN,RL  & \cite{RLGCNlogic} \\ \cmidrule(lr){2-4}
    & To generate the optimal synthesis flows. & RL      & \cite{RLlogic}     \\ \cmidrule(lr){1-4}

    \multirow{2}{2.5cm}{\centering Placement}
    & To train, predict, and evaluate potential datapaths. & SVM,NN & \cite{pade_placer} \\ \cmidrule(lr){2-4}
    & To make placement decisions. & GCN,RL & \cite{mirhoseini2020chip} \\ \cmidrule(lr){1-4}

    \multirow{7}{2.5cm}{\centering Routing}
    & To detect DRC hotspot and DRV count. & CNN & \cite{Routenet}\\ \cmidrule(lr){2-4} 
    & \multirow{5}{6cm}{To predict routing congestion.}
       & CNN & \cite{J-net} \\ \cmidrule(lr){3-4}
    &  & GAN & \cite{9045178} \\ \cmidrule(lr){3-4}
    &  & ML  & \cite{8533535} \\ \cmidrule(lr){3-4}
    &  & MARS & \cite{ML4DR} \\ \cmidrule(lr){2-4}
    & To predict routability of a given placement. & MARS,SVM & \cite{ICCD2016} \\ \cmidrule(lr){2-4}
    & To model on-chip router performance. & MARS & \cite{Jeong} \\ \cmidrule(lr){2-4}
    & To predict wirelength. & LDA, KNN & \cite{8715016} \\ \cmidrule(lr){2-4}
    & To predict the circuit performance after placement stage. & ML & \cite{8394712} \\ \cmidrule(lr){2-4}
    & To predict detailed routing result after global routing. & ML & \cite{7835438} \\ \cmidrule(lr){2-4}
    & \multirow{2}{6cm}{To model sign-off timing analysis.}
       & RF & \cite{8807063} \\ \cmidrule(lr){3-4}
    &  & LR & \cite{6681682} \\ \cmidrule(lr){2-4}
    & To predict and optimize the clock tree. & GCN,CNN,RL & \cite{GAN-CTS} \\ \cmidrule(lr){1-4}


    \multirow{5}{2.5cm}{\centering Power Deliver Network Synthesis and IR Drop Predictions}
    & To predict incremental static IR drop. & XGBoost & \cite{IncPIRD} \\ \cmidrule(lr){2-4}
    & To predict dynamic IR drop by regressing. & CNN & \cite{PowerNet} \\ \cmidrule(lr){2-4}
    & To predict electromigration-induced IR drop. & GAN & \cite{GridNet} \\ \cmidrule(lr){2-4}
    & To choose the power grid template. & MLP,CNN & \cite{Template} \\ \cmidrule(lr){2-4}
    & To predict bump inductance. & SVM,ANN,MARS & \cite{10.1145/3287624.3287689}  \\ \cmidrule(lr){1-4}

    \multirow{4}{2.5cm}{\centering 3D Integration}
    & To advance the tier partition. & GNN & \cite{TPGNN} \\ \cmidrule(lr){2-4}
    & To model the path delay variation. & MARS & \cite{samal2016machine} \\ \cmidrule(lr){2-4}
    & \multirow{2}{2.5cm}{To optimize 3D designs.}
        & Local Search & \cite{7372639} \\ \cmidrule(lr){3-4}
    &   & BO & \cite{7850943} \\ \cmidrule(lr){1-4}

    \multirow{2}{2.5cm}{\centering Other}
    & To predict the embedded memory timing failure. & ML & \cite{7428008} \\ \cmidrule(lr){2-4}
    & To predict aging effect. & RF & \cite{8060438} \\ 

\bottomrule
    \end{tabular}
\end{table}
\subsection{Design Challenges for 3D Integration}  \label{sec:physical:3d}



3D integration is gaining more attention as a promising approach to further improve the integration density. It has been widely applied in memory fabrication by stacking memory over logic.

Different from the 2D design, 3D integration introduces die-to-die variation, which does not exist in 2D modeling.
The data or clock path may cross different dies in through-silicon via (TSV)-based 3D IC.
Therefore, the conventional variation modeling methods, such as on-chip variation (OCV), advanced OCV (AOCV), parametric OCV (POCV),
are not able to accurately capture the path delay~\cite{samal2016machine}. \citet{samal2016machine} use MARS to model the path delay variation in 3D ICs.

3D integration also brings challenges to the design optimization due to the expanded design space and the overhead of design evaluation.
To tackle these challenges, several studies~\cite{7372639,7850943,samal2016machine} have utilized design space exploration methods based on machine learning to facilitate 3D integration optimization.

The state-of-the-art 3D placement methods~\cite{Compact-2D,S2D} perform bin-based tier partitioning on 2D placement and routing design.
However, the bin-based partitioning can cause significant quality degradation to the 3D design because of the unawareness of the design hierarchy and technology.
Considering the graph-like nature of the VLSI circuits, \citet{TPGNN} proposed a GNN-based unsupervised framework (TP-GNN) for tier partitioning.
TP-GNN first performs the hierarchy-aware edge contraction to acquire the clique-based graph where nodes within the same hierarchy can be contracted into supernodes.
Moreover, the hierarchy and the timing information is included in the initial feature of each node before GNN training.
Then the unsupervised GNN learning can be applied to general 3D design.
After the GNN training, the weighted k-means clustering is performed on the clique-based graph for the tier assignment based on the learned representation.
The proposed TP-GNN framework is validated on experiments of RISC-V based multi-core system and NETCARD from ISPD 2012 benchmark.
The experiment results indicate 7.7\% better wirelength, 27.4\% higher effective frequency and 20.3\% performance improvement.




\subsection{Other Predictions}\label{sec:physical:other}

For other parameters, \citet{7428008} adopt HSM to predict the embedded memory timing failure during initial floorplan design. \citet{8060438} work on aging effect prediction for high-dimensional correlated on-chip variations using random forest.

\subsection{Summary of Machine Learning for Logic Synthesis and Physical Design}
We summarize recent studies on ML for logic synthesis and physical design in \Cref{tb::physical_summary}.
For logic synthesis, researchers focus on predicting and evaluating the optimal synthesis flows.
Currently, these studies optimize the synthesis flow based on the primitives of existing tools.
In the future, we expect to see more advanced algorithms for logic synthesis be explored, and more metrics can be formulated to evaluate the results of logic synthesis.
Besides, applying machine learning to logic synthesis for emerging technologies is also an interesting direction.

In the physical design stage, recent studies mainly aim to improve the efficiency and accuracy by predicting the related information that traditionally needs further simulation. A popular practice is to formulate the EDA task as a computer vision (CV) task.
In the future, we expect to see more studies that incorporate advanced techniques (e.g., neural architecture search, automatic feature generation, unsupervised learning) to achieve better routing and placement results.


\section{Lithography and Mask synthesis}
\label{sec:mask}


Lithography is a key step in semiconductor manufacturing, which turns the designed circuit and layout into real objects. 
{Two popular research directions are lithography hotspot detection and mask optimization.
To improve yield, lithography hotspot detection is introduced after the physical implementation flow to identify process-sensitive patterns prior to the manufacturing. The complete optical simulation is always time-consuming, so it is necessary to analyze the routed layout by machine learning to reduce lithography hotspots in the early stages.
Mask optimization tries to compensate diffraction information loss of design patterns such that the remaining pattern after lithography is as close to the design patterns as possible.}
{Mask optimization plays an important role in VLSI design and fabrication flow, which is a very complicated optimization problem with high verification costs caused by expensive lithography simulation. Unlike the hotspot detection studies in \Cref{sec:lithography_hotspot} that take placement \& routing stages into consideration, mask optimization focuses only on the lithography process, ensuring that the fabricated chip matches the designed layout. Optical proximity correction (OPC) and sub-resolution assist feature (SRAF) insertion are two main methods to optimize the mask and improve the printability of the target pattern.}

\subsection{Lithography Hotspot Detection}\label{sec:lithography_hotspot}

For lithography hotspot detection, {\citet{AENEID} uses} SVM for hotspot detection and small neural network for routing path prediction on each grid.
To achieve better feature representation, {\citet{8360060} introduces} feature tensor extraction, which is aware of the spatial relations of layout patterns.
{This} work develops a batch-biased learning algorithm, which provides better trade-offs between accuracy and false alarms.
Besides, there are also attempts to check inter-layer failures with deep learning solutions.
A representative solution is {proposed by \citet{yang2019detecting}}.
{They employ} an adaptive squish layout representation for efficient metal-to-via failure check. 
{Different layout-friendly neural network architectures are also investigated these include vanilla VGG \cite{yang2017imbalance}, shallow CNN \cite{8360060} and binary ResNet \cite{jiang2020efficient}. }

\begin{figure}
	\centering
	\subfigure[Traditional Hotspot Detection] {\includegraphics[width=.56\textwidth]{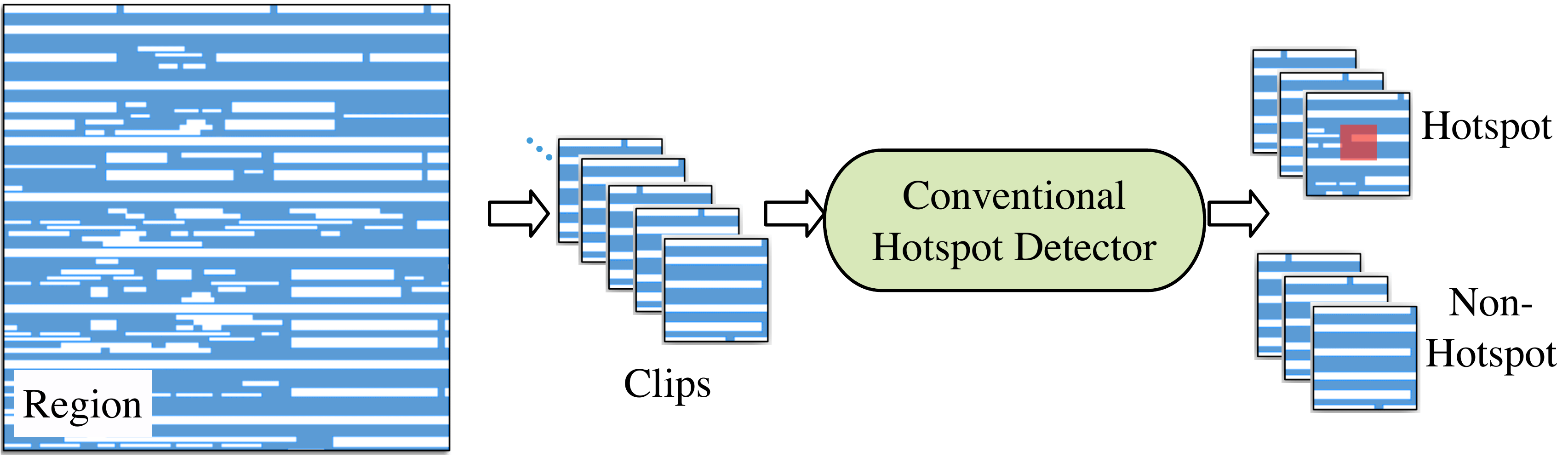}}
	\subfigure[Region-based Hotspot Detection]{\includegraphics[width=.56\textwidth]{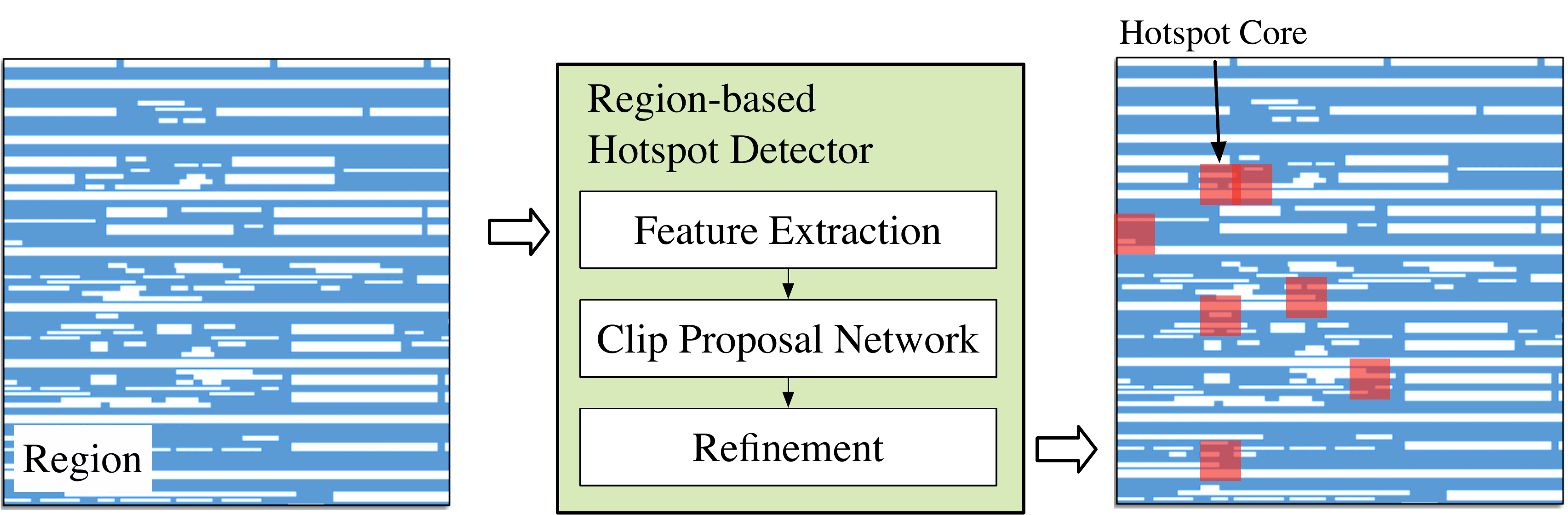}}
    \caption{Region-based hotspot detection promises better performance (reproduced from~\cite{chen2020faster}).}
	\label{fig:chen2020faster}
\end{figure}

With the increased chip complexity, traditional deep learning/machine learning-based solutions are facing challenges from both runtime and detection accuracy.
{\citet{chen2020faster}} recently propose an end-to-end trainable object detection model for large scale hotspot detection.
The framework takes the input of a full/large-scale layout design and {localizes} the area that hotspots might {occur} (see \Cref{fig:chen2020faster}). 
In \cite{lithattention}, an attention-based CNN {with} inception-based backbone is developed for better feature embeddings.

\begin{figure}[htbp]
    \centering
    \includegraphics[width=0.518\linewidth]{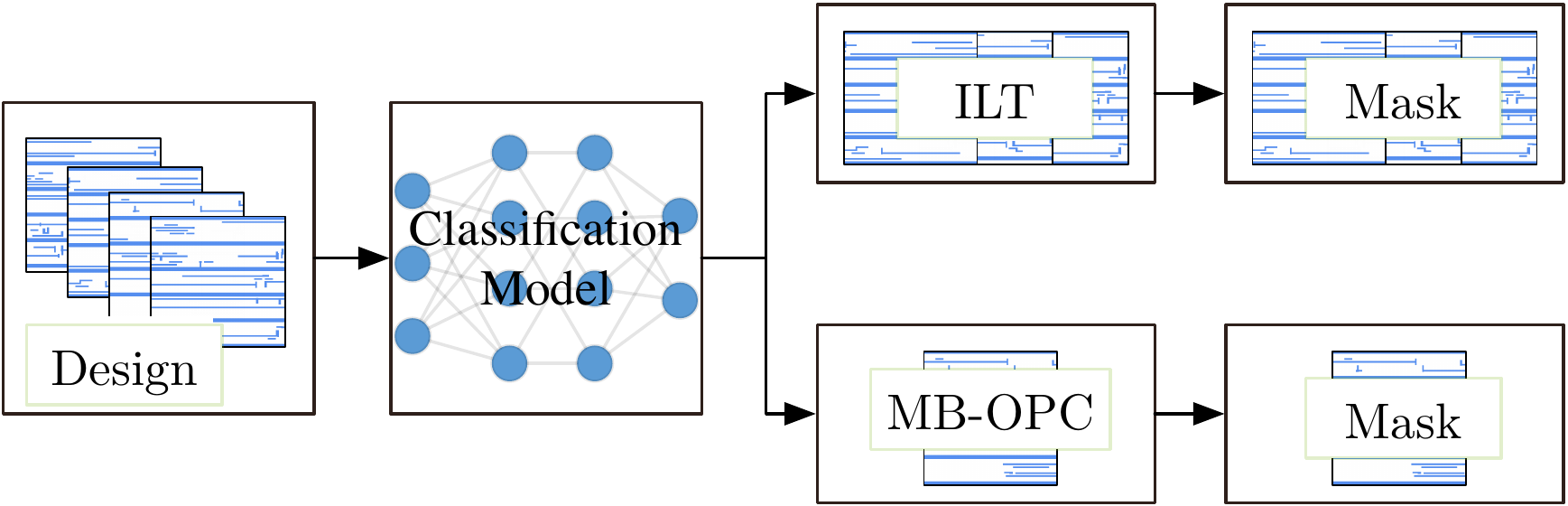}
    \caption{A heterogeneous OPC framework (reproduced from \cite{mask-optimization}).}
    \label{fig:hetro-opc}
\end{figure}

{\subsection{Machine Learning for Optical Proximity Correction}} \label{sec:mask:opc}
For OPC, inverse lithography technique (ILT) and model-based OPC are two representative mask optimization methodologies, and each of which has its own advantages and disadvantages.
\citet{mask-optimization} propose a {heterogeneous} OPC framework that assists mask layout optimization, where a deterministic ML model is built to choose the appropriate one from multiple OPC solutions for a given design, as shown in \Cref{fig:hetro-opc}.

\begin{figure}[htbp]
    \centering
    \includegraphics[width=0.66\linewidth]{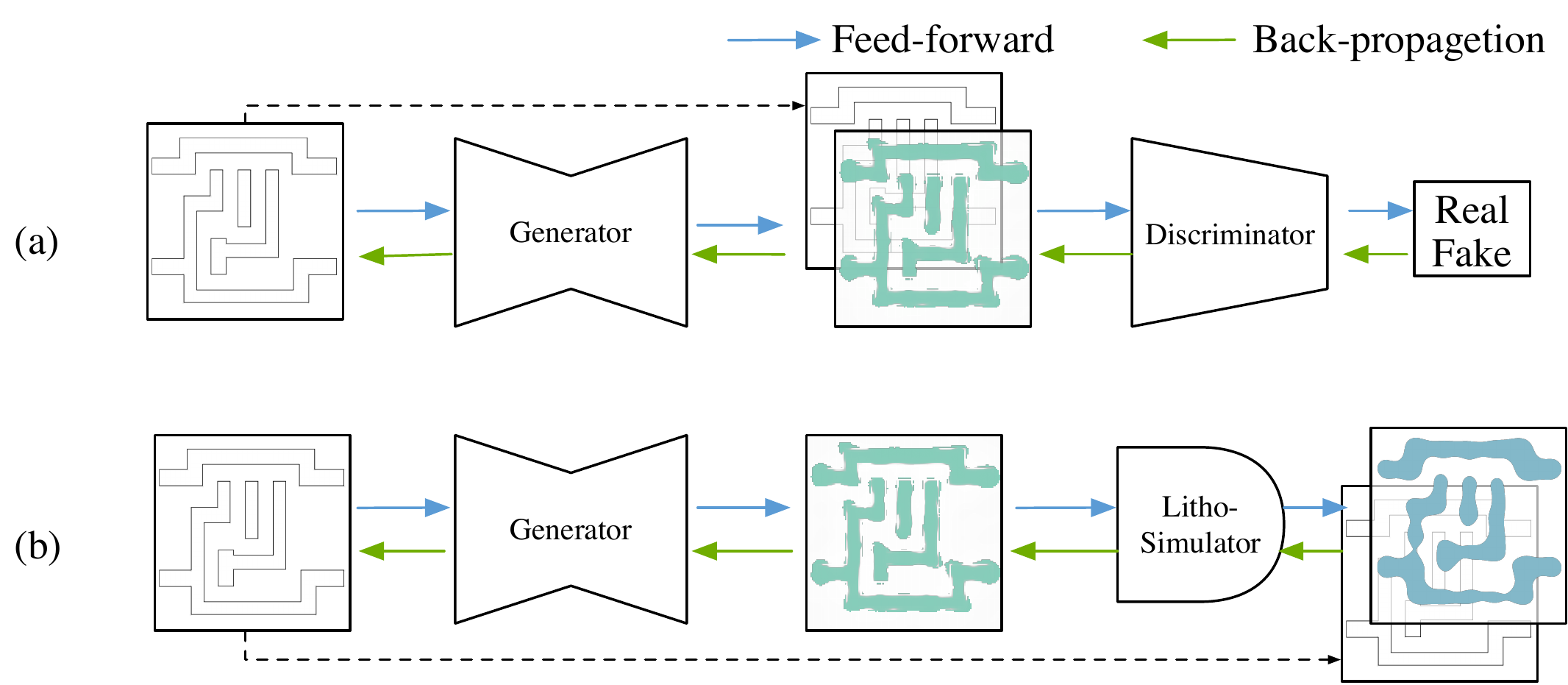}
    \caption{The training flow of (a) GAN-OPC and (b) ILT-Guided Pre-training (reproduced from~\cite{yang2018gan}).}
    \label{gan-opc-test}
\end{figure}

With the improvement of semiconductor technology and the scaling down of ICs, {traditional} OPC methodologies are becoming more and more complicated and time-consuming.
\citet{yang2018gan} propose a new OPC method based on generative adversarial network (GAN).
A Generator (G) is used to generate the mask pattern from the target pattern, and a discriminator (D) is used to estimate the quality of the generated mask.
GAN-OPC can avoid complicated computation in ILT-based OPC, but it faces the problem that the algorithm is hard to converge.
To deal with this problem, ILT-guided pre-training is proposed.
In the pre-training stage, the D network is replaced with the ILT convolution model, and only {the} G network is trained.
After pre-training, {the} ILT model that has huge cost is removed, and the whole GAN is trained.
The training flow of GAN-OPC and ILT-guided pre-training is shown in \Cref{gan-opc-test}.
The experimental results show that the GAN-based methodology can accelerate ILT based OPC significantly and generate more accurate mask patterns.

{Traditional ILT-based OPC methods are costly and result in highly complex masks where many rectangular variable-shaped-beam (VSB) shots exist.
To solve this problem, \citet{9256592} {propose} an ML-based OPC algorithm named neural-ILT, which uses a neural network to replace the costly ILT process. The loss function is specially designed to reduce the mask complexity, which gives punishment to complicated output mask patterns. In addition, for fast litho-simulation, a CUDA-based accelerator is proposed as well, which can save 96\% simulation time. The experimental results show that neural-ILT achieves a 70$\times$ speedup and 0.43$\times$ mask complexity compared with traditional ILT methods.}

Recently, \citet{DAMO} propose DAMO, an end-to-end OPC framework to tackle the full-chip scale. The lithography simulator and mask generator share the same deep conditional GAN (DCGAN), which {is dedicatedly designed and can provide a competitively high resolution}. The proposed DCGAN {adopts} UNet++~\cite{UNet++} backbone and {adds} residual blocks at the bottleneck of UNet++.
To further apply DAMO on full-chip layouts, a coarse-to-fine window splitting algorithm is proposed.
First, it locates the regions of high via density and then runs KMeans++ algorithm on each cluster containing the via pattern to find the best splitting window.
Results on ISPD 2019 full-chip layout show that DAMO outperforms state-of-the-art OPC solutions in both academia~\cite{sarf-insertion} and an industrial toolkit.


\subsection{Machine Learning for SRAF Insertion} \label{sec:mask:sraf}

\begin{figure}[htbp]
    \centering
    \includegraphics[width=0.58\linewidth]{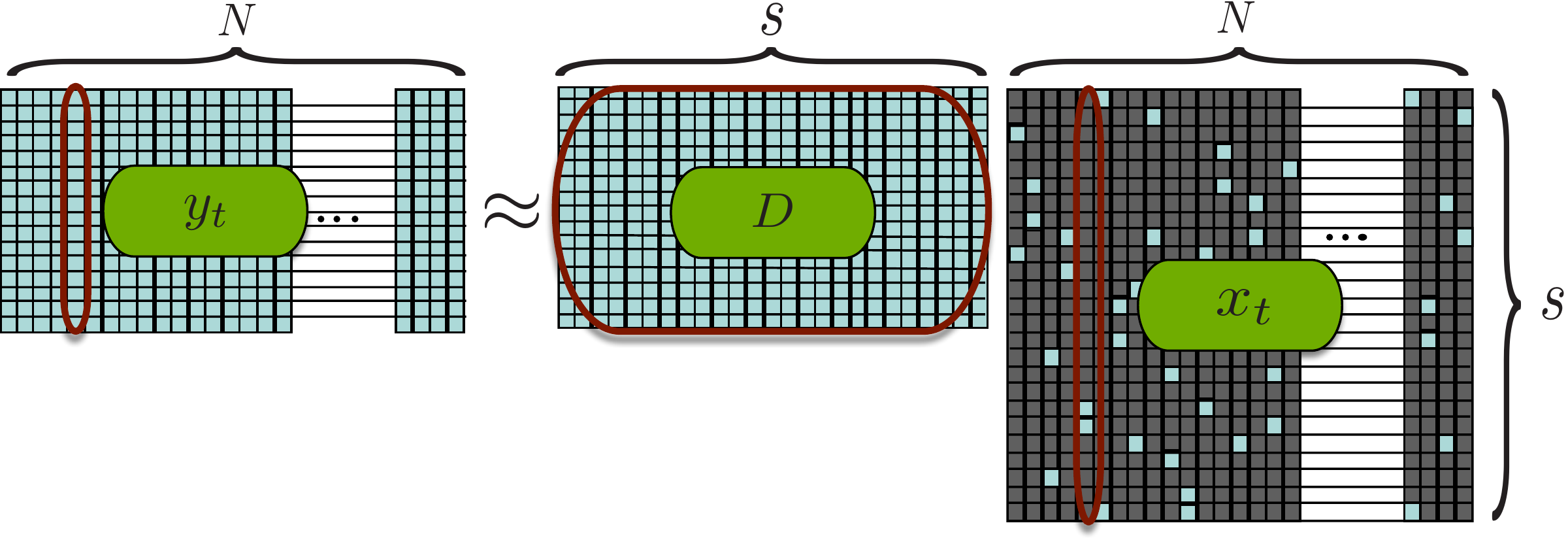}
    \caption{Dictionary learning based feature extraction (reproduced from~\cite{sarf-insertion}).}
    \label{fig:sraf-mapping}
\end{figure}

Several studies have investigated ML-aided SRAF insertion techniques. \citet{Xu2016AML} propose an SRAF insertion framework based on ML techniques. \citet{sarf-insertion} propose a framework with a better feature extraction strategy. \Cref{fig:sraf-mapping} shows the feature extraction stage. After their concentric circle area sampling (CCAS) method, high-dimension features $x_t$ are mapped into a discriminative low-dimension features $y_t$ through dictionary training by multiplication of an atom matrix $D$.
The atom matrix is the dictionary consists of representative atoms of the original features.
Then, the sparse codes $y_t$ are used as the input of a machine learning model, more specifically,
a logistic regression model that outputs a probability map indicating whether SRAF should be inserted at each grid.
Then, the authors formulate and solve the SRAF insertion problem as an integer linear programming based on the probability grid and various SRAF design rules.

\subsection{Machine Learning for Lithography Simulation} \label{sec:mask:lithosimulation}

\begin{figure}[htb!]
    \centering
    \includegraphics[width=0.58\linewidth]{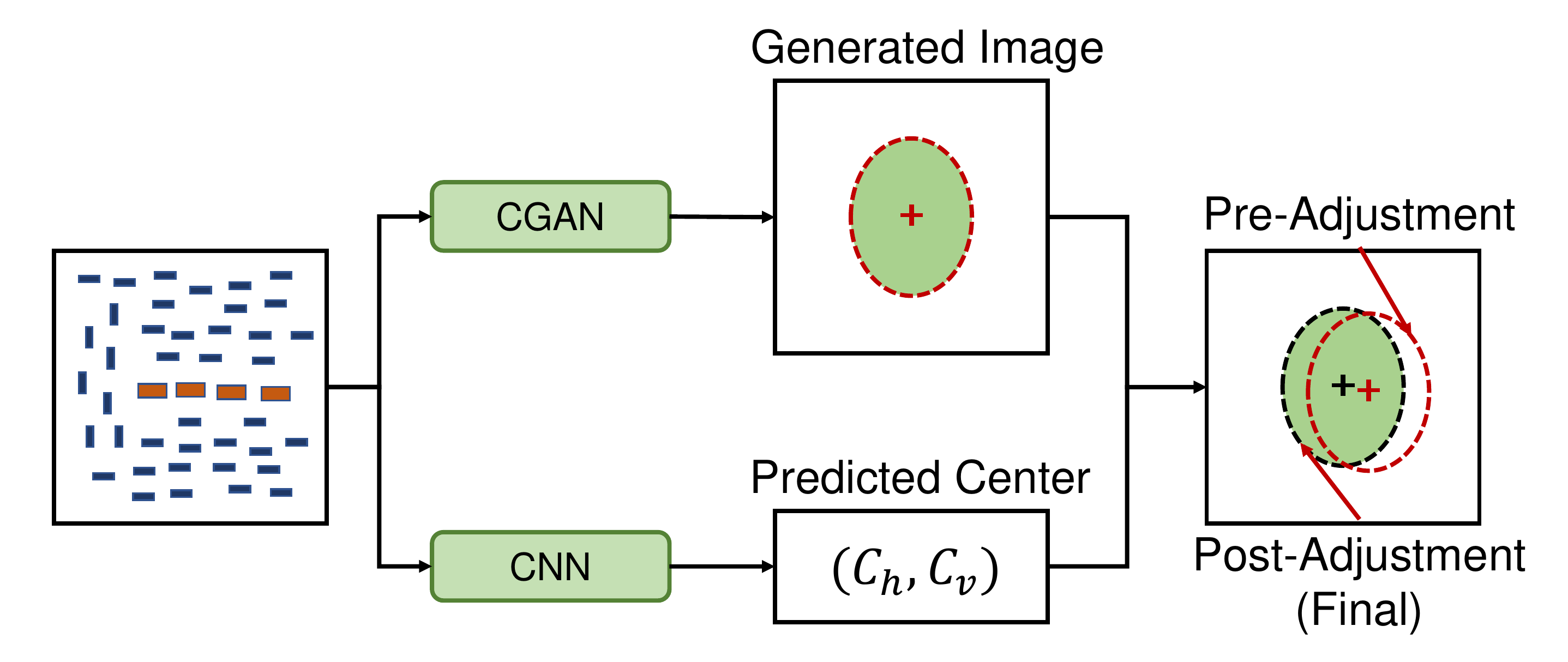}
    \caption{The LithoGAN framework (reproduced from~\cite{ye2019lithogan}).}
    \label{test_fig6}
\end{figure}

There are also studies that focus on fast simulation of the tedious lithography process.
Traditional lithography simulation contains multiple steps, such as optical model building, resist model building, and resist pattern generation.
{LithoGAN~\cite{ye2019lithogan} proposes} an end-to-end lithography modeling method by using GAN, of which the framework is shown in \Cref{test_fig6}.
Specifically, a conditional GAN is trained to map the mask pattern to a resist pattern. However, due to the characteristic of GAN, the generated shape pattern is good, while the position of the pattern is not precise. To tackle this problem, LithoGAN adopts a conditional GAN for shape modeling and a CNN for center prediction. The experimental results show that LithoGAN can predict the resist pattern with high accuracy, {and} this algorithm can reduce the lithography simulation time for several orders of magnitude.
\cite{DAMO} is also equipped with a machine learning-based lithography simulator that can output via contours accurately to assist via-oriented OPC.


{\subsection{Summary}
This section reviews ML techniques used in the design for manufacturability stage that include lithography hotspot detection, mask optimization and lithography modeling.
Related studies are summarized in \Cref{mask_summary}.}

\begin{table*}
    \caption{{Summary of ML for lithography and mask optimization} }
    \label{mask_summary}
    \begin{center}
    	\footnotesize
    	{
\begin{tabular}{p{2.5cm}|p{6cm}|p{2cm}|p{1.6cm}}
\toprule
     \textbf{\centering{Task}} & \textbf{Work} & \textbf{ML Algorithm} & \textbf{References}\\ \midrule
         \multirow{5}{2.5cm}{\centering Lithography Hotspot Detection}
     & \multirow{2}{6cm}{To detect single layer layout lithography hotspots.}
     & SVM, NN & \cite{AENEID} \\ \cmidrule(lr){3-4}
     &   & CNN & \cite{8360060,yang2017imbalance,jiang2020efficient} \\ \cmidrule(lr){2-4}
     & To detect multilayer layout lithography hotspots.  & CNN & \cite{yang2019detecting} \\ \cmidrule(lr){2-4}
     & \multirow{2}{6cm}{To fast detect large scale lithography hotspots.}
     & CNN & \cite{chen2020faster} \\ \cmidrule(lr){3-4}
     &   & Attention & \cite{lithattention} \\ \cmidrule(lr){1-4}
     \multirow{6}{3cm}{\centering OPC}
     & Heterogeneous OPC & CNN & \cite{mask-optimization} \\ \cmidrule(lr){2-4} 
     & GAN-OPC & GAN & \cite{yang2018gan}\\ \cmidrule(lr){2-4}
     & Neural ILT  & CNN & \cite{9256592} \\ \cmidrule(lr){2-4}
     & DAMO & DCGAN & \cite{DAMO} \\ \cmidrule(lr){1-4}
     \multirow{5}{3cm}{\centering SRAF insertion}& ML-based SRAF generation & Decision Tree, Regression & \cite{Xu2016AML}\\ \cmidrule(lr){2-4}
     & SRAF insertion & Dictionary learning & \cite{sarf-insertion}\\ \cmidrule{1-4}
     \multirow{2}{3cm}{\centering Litho-simulation} & LithoGAN & CGAN, CNN & \cite{ye2019lithogan} \\ \cmidrule(lr){2-4}
      & DAMO & DCGAN & \cite{DAMO} \\
\bottomrule
    \end{tabular}}
    \end{center}
\end{table*}

\section{Analog Design}
\label{sec:analog}
Despite the promotion of digital circuits, the analog counterpart is still irreplaceable in applications like nature signal processing, high speed I/O and drive electronics~\cite{razavi}. Unlike digital circuit design, analog design demands lots of manual work and expert knowledge, which often makes it the bottleneck of the job. For example, the analog/digital converter and {Radio Frequency (RF)\footnote{With a slight abuse of acronym, RF stands for both Random Forest, and Radio Frequency. The meaning should be clear from the context.}} transceiver only occupy a small fraction of area but cost the majority of design efforts in a typical mixed-signal {System-on-Chip (SoC)}, compared to other digital processors~\cite{nuo}. 

The reason for the discrepancy can be summarized as follows: 
{1) Analog circuits have a larger design space in terms of device size and topology than digital circuits. Sophisticated efforts are required to achieve satisfactory results. 2) The specifications of analog design are variable for different applications. It is difficult to construct a uniform framework to evaluate and optimize different analog designs. 3) Analog signals are more susceptible to noise and process-voltage-temperature variations, which cost additional efforts in validation and verification.}

\subsection{The Design Flow of Analog Circuits}

\begin{figure}
\centering
\includegraphics[width=.68\linewidth]{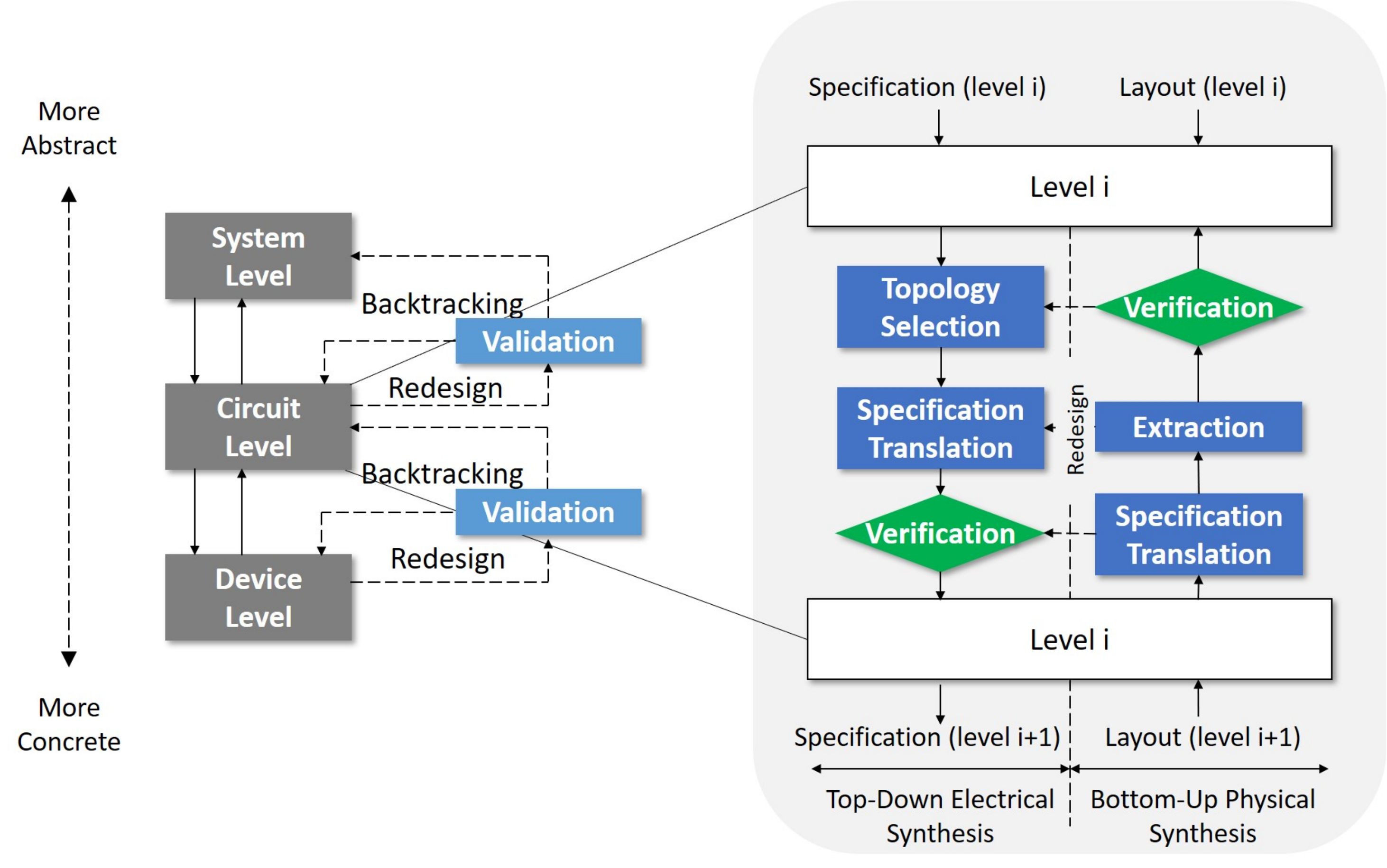}
\caption{Hierarchical levels of analog design flow (reproduced from~\cite{nuo}).}
\label{flow}
\end{figure}

\citet{Gielen} provide the design flow followed by most analog designers. As shown in \Cref{flow}, it includes both top-down design steps from system level to device-level optimizations and bottom-up layout synthesis and verification. In the top-down flow, designers choose proper topology, which satisfies system specifications in the circuit level. Then device sizes are optimized in the device level.
{The topology design and device sizing constitute the pre-layout design. After the schematic is well-designed, designers draw the layout of the circuit. Then they extract parasitics from the layout and simulate the circuit with parasitics. This is known as post-layout simulations. If the post-layout simulation fails to satisfy the specifications, designers need to resize the parameters and repeat the process again. This process can go for many iterations before the layout is done~\cite{shook2020mlparest}.}
 
Although analog design automation has improved significantly over the past few decades, automatic tools cannot replace manual work in the design flow~\cite{review} yet. 
Recently, researchers are trying to introduce machine learning techniques to solve analog design problems. Their attempts range from topology selection at the circuit level to device sizing at the device level {as well as the analog layout in the physical level}.

\subsection{Machine Learning for Circuit Topology Design Automation}

Typically, topology design is the first step of analog circuits design, followed by the determination of device sizes and parameters. The process is time-consuming, and unsuitable topology will lead to redesign from the very beginning. Traditionally, topology design relies on the knowledge and experiences of expert designers.
As the scale and demand of analog circuits are increasing, CAD tools are urgently needed by engineers.
Despite this, automation tools for topology design are still much less explored due to its high degree of freedom.

Researchers have attempted to use ML methods to speed up the design process. Some researchers~\cite{fasy_tcad1996,fuzzy_,inference_sisa2018} deal with topology selection problem, selecting the most suitable topology from several available candidate. \citet{analog_edaa2016} focus on extracting well-known building blocks in circuit topology. Recently, \citet{electric_arxiv2020} use RNN and hypernetwork to generate two-port circuit topology.

\subsubsection{Topology Selection} \label{sec:analog:topologyselection}
For common-used circuit functional units, like amplifiers, designers may not need to design from the beginning. Instead, it is possible to choose from a fixed set of available alternatives. It is a much more simple problem than designing from {scratch.} 
Early in 1996, \citet{fasy_tcad1996,fuzzy_} put forward a fuzzy-logic based topology selection tool called FASY. They use fuzzy logic to describe relationships between specifications (e.g., DC gain) and alternatives and use backpropagation to train the optimizer. 
More recent research~\cite{inference_sisa2018} uses CNN as the classifier. They train CNN with circuit specifications as the inputs and the topology indexes as the labels. 

The main problem with the topology selection methods is that the data collection and the training procedure are time-consuming. Therefore, topology selection is efficient only when repetitive designs are needed such that a trained model can be reused. 

\subsubsection{Topological Feature Extraction}
One challenge of topology design automation is to make algorithms learn the complex relationships between components. To make these relationships more understandable, researchers focus on defining and extracting features from circuit topology. \citet{analog_edaa2016} present algorithms for both supervised feature extraction and unsupervised learning of new connections between known building blocks. 
The algorithms are also designed to find hierarchical structures, isolate generic templates (patterns), and recognize overlaps among structures.
{Symmetry constraint are one of the most essential topoligical features in circuits. \citet{liu2020s3det} propose a spectral analysis method to detect system symmetry with graph similarity. With a graph representation of circuits, their method is capable of handling passive devices as well. \citet{kunal2020general} propose a GNN-based methodology for automated generation of symmetry constraints. It can hierarchically detect symmetry constraints in multiple levels and works well in a variety of circuit designs.}

\subsubsection{Topology Generation}


The aforementioned studies do not directly generate a topology. A recent study~\cite{electric_arxiv2020} makes the first attempt to generate circuit topology for given specifications. Their focus is limited to two-port circuits.
They utilize an RNN and Hypernetwork to solve the topology generation problem and report better performance than the traditional methods when the {inductor} circuit length $n \ge 4$.

\subsection{Machine Learning for Device Sizing Automation} \label{sec:analog:devicesizing}
\subsubsection{Reinforcement Learning Based Device Sizing}
The problem of device sizing can be formulated as follows: 
\begin{alignat}{2}
\arg\min_x \quad & \sum_x q_c(x),\nonumber\\
\mbox{s.t.}\quad
&f_h(x)\geq y_h,
\end{alignat}
where $x\in\mathbb{R}^n$ denotes the design parameters, including the size of each transistors, capacitors and resistors. $y\in\mathbb{R}^m$ denotes the specifications, including the rigid targets $y_h\in\mathbb{R}^{m_1}$ such as bandwidths, DC gains or phase margins and the optimization targets $y_o\in\mathbb{R}^{m_2}$ such as power or area. The simulator $f$ is defined as the map from parameters to specifications. To normalize the contribution of different specifications, the objective function is defined as $q_c(x)=f_o(x)/y_o$.

\begin{figure*}
    \centering
    \includegraphics[width=.56\linewidth]{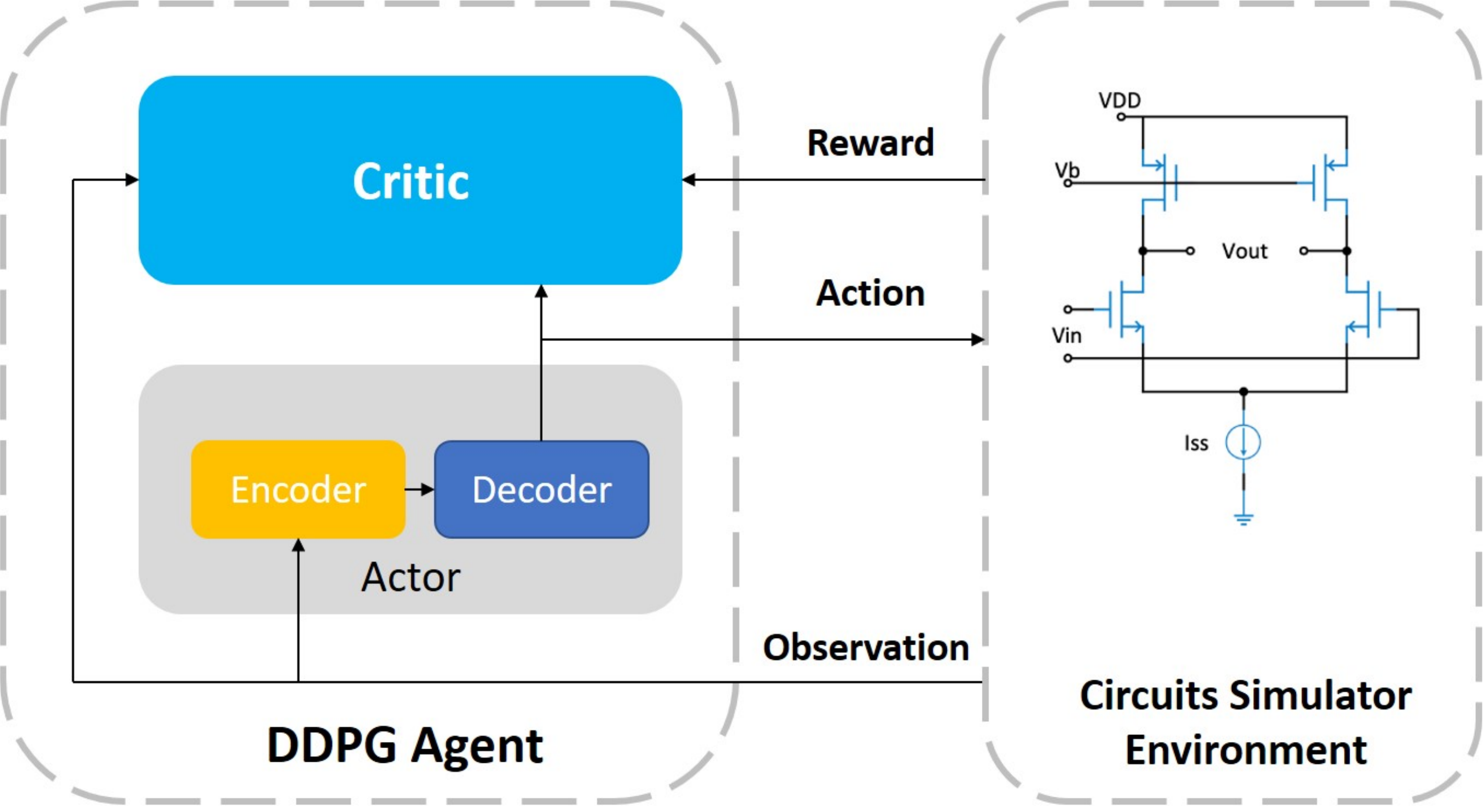}
    \caption{The framework of reinforcement learning for device sizing (reproduced from~\cite{learning}).}
    \label{ro}
\end{figure*}

Based on this optimization model, \citet{learning} apply the reinforcement learning technique to deal with device sizing problems.
\Cref{ro} illustrates the proposed reinforcement learning framework. At each environment step, the observations from the simulator are fed to the agent. A reward is calculated by the value network based on current performance. Then, the agent responds with an action to update the device sizes. Because the transistors are both affected by their local status {(e.g., transconductance $g_m$, drain current $I_{ds}$, etc.)} and the global status (DC operating points) of the circuit, the optimization of each transistor is not independent. To promote learning performance and efficiency, the authors use a multi-step environment, where the agent receives both the local status of the corresponding transistor and the global status.

Although the device sizing problem is automated by the reinforcement learning approach, the training process depends heavily on efficient simulation tools. However, current simulation tools can only satisfy the need for schematic simulations. As for post-layout simulation that requires parasitic extraction, the time of each training iteration increases significantly. To reduce the simulation overhead, \citet{autockt} introduce transfer learning techniques into reinforcement learning. In the proposed approach, the agent is trained by schematic simulations and validated by post-layout simulations. The authors show that, with some additional iterations on deployment, the proposed approach can bring {9.4$\times$} acceleration compared to previous approaches.


Following their previous work~\cite{learning}, the authors utilize GCN to enhance the transferability of reinforcement learning methods~\cite{gcn-rl}. Unlike traditional multi-layer agents, the GCN-based agent
extracts topology information from circuit netlists. In a GCN layer, each transistor is represented by a hidden neuron calculated by aggregating feature vectors from its neighbors. Specifically, the calculation can be written as:

\begin{equation}
H^{(l+1)} = \sigma(\tilde{D}^{-1/2}\tilde{A}\tilde{D}^{-1/2})H^{(l)}W^{(l)},
\end{equation}
where $\tilde{A}$ is the adjacency matrix $A$ of the circuit topology plus the identity matrix $I_N$. $\tilde{D}_{ii} = \sum\tilde{A_{ij}}$ is a diagonal matrix. And $H^{(l+1)}$ is the hidden features of the $l$th layer. The weight matrix $W^{(l)}$ is a trainable matrix updated by Deep Deterministic Policy Gradient (DDPG)~\cite{ddpg}. Because different circuits with the same function have similar design principles (e.g., two-stage and three-stage amplifier). The weight matrix trained for one circuit can be reused by another circuit. Besides the topologies transferring, the GCN-based RL agent is able to port existing designs from one technology node to another by sharing the weight matrices.




\subsubsection{Artificial Neural Network Based Device Sizing}
\citet{nuo} propose a data augmentation method to increase the generalization ability of the trained ML model. Specifically, the original training set $T$ is replaced by augmented set $T'=T\cup T_1 \cup T_2 ... \cup T_k$. For each $T_i$, the relationship between its sample $x'_i$ and original sample $x_i$ is formulated as follows:
\begin{equation}
x'_i=x_i+(\frac{\gamma}{M}\sum_{j=1}^{M}x_j)\Delta\Gamma,
\end{equation}
where $\gamma\in [0,1]$ is a hyper-parameter used to adjust the mean value. And $\Delta$ and $\Gamma$ denote a diagonal matrix composed by random value of [0,1] and value in ${-1,1}$, respectively. For $spec_i$ to maximize like DC gain, $\Gamma_i$ takes value -1. Conversely, it takes value 1 for $spec_i$ to minimize, like power or area. As a result, $K$ copies with worse specifications are generated for each sample $x_i$. {The models trained on the augmented dataset are more robust.} 


Besides the augmentation method, this paper proposes to use a 3-layer MLP model to conduct regression and classification. 
Given circuit performances as the input, the model outputs circuit information as two parts: 1) the size of devices in multiple topologies; 2) the classification of different topologies. The device sizing problem is solved by regression, while the topology selection is solved by classification. Compared to the simple regression models, the regression-and-classification model obtains the best performance.

\subsubsection{Machine Learning Based Prediction Methods}
As mentioned above, the time cost by simulation is the main overhead of training models, especially for post-layout simulation. In order to speed up the training process, \citet{bag} and \citet{svm} use DNN and SVM to predict the simulator. 
\citet{bag} use the device information of two circuits as the input of the DNN predictor.
The model outputs the relative superiority of the two circuits on each specification instead of the absolute values.
Because the {prediction problem is non-convex and even ill-posed}, and the training data is also limited by computational resources. Learning to compare (predict the superiority) is a relatively easy task compared to directly fitting each specification. Besides, enumerating each pair of circuit designs enlarge the training set by {$\frac{N}{2}\times$, where $N$ denotes the number of circuit designs.}


\subsubsection{Comparison and Discussion on Device Sizing}
\begin{table} 
    \centering
    \caption{Comparison of different device sizing methods} 
    \label{dt}
    \footnotesize
        \begin{tabular}{p{4cm}|p{2.5cm}|p{3.5cm}|p{1cm}}
            \toprule
            \textbf{ML Algorithm} & \textbf{Simulation tools} & \textbf{Nums of simulation (Two-Stage OP)} & \textbf{Reference} \\    \midrule
           Reinforcement learning &  Commercial tools & 1e4  &  \cite{learning} \\ \midrule
            Genetic algorithm & DNN & 1e2 & \cite{bag}   \\ \midrule
            Reinforcement learning+Transfer learning & Commercial tools & 1e4/10 (training/deployment) & \cite{autockt} \\\midrule
            Reinforcement learning+GCN & Commercial tools & 1e4/100 (training/deployment) & \cite{gcn-rl} \\ \midrule
            ANN & Commercial tools & 1e4/1 (training/deployment) & \cite{nuo} \\ \midrule
            Genetic algorithm & SVM & 1e2 & \cite{svm} \\ \bottomrule
        \end{tabular} 
\end{table}

\Cref{dt} lists introduced methods and their performance. The widely-studied two-stage operational amplifier (OPA)  is adopted as an example for comparison.
Instead of performance, sample efficiency is used as the criterion because two-stage OPA is a relatively simple design and different algorithms can achieve comparable circuit performance.
It is shown that machine learning algorithms require more simulations on the training phase than traditional genetic methods. But only a few iterations of inference are needed when deploying the model. Thus, ML-based methods have more potential in large scale applications at the cost of increased training costs. On the other hand, genetic algorithms combined with ML-based predictor is a popular solution to reduce the number of needed simulations.
Note that different learning algorithms have been adequately verified on simple designs like two-stage OPA. However, designing complicated circuits is still challenging. 

\subsection{Machine Learning for Analog Layout}


\begin{table*} 
    \caption{{Summary of ML for analog layout} }
    \label{tab:layout}
    \begin{center} 
    	\footnotesize
\begin{tabular}{p{3cm}|p{3cm}|p{3cm}|p{3cm}} 
\toprule
    \textbf{\centering{Stage}} & \textbf{Task} & \textbf{ML Algorithm} & \textbf{References}\\ \midrule
    \multirow{3}{3cm}{\centering Pre-layout Preparation}
     & circuit hierarchy generation & GCN & \cite{kunal2020gana} \\ \cmidrule(lr){2-4}
     & \multirow{2}{3cm}{parasitics estimation}
     & GNN & \cite{ren2020paragraph} \\ \cmidrule(lr){3-4}
     &  & Random Forest & \cite{shook2020mlparest}\\ \midrule
    \multirow{3}{3cm}{\centering Layout Generation}
     & well generation & GAN & \cite{xu2019wellgan} \\ \cmidrule(lr){2-4}
     & closed-loop layout synthesis & Bayesian Optimization & \cite{liu2020closing} \\ \cmidrule(lr){2-4}
     & routing & VAE & \cite{zhu2019genius} \\ \midrule
    \multirow{4}{3cm}{\centering Post-layer Evaluation}
     & electromagnetic properties estimation & GNN & \cite{zhang2019circuit} \\ \cmidrule(lr){2-4}
     & \multirow{3}{3cm}{performance prediction}
     & SVM, random forest, NN & \cite{li2020exploring}\\ \cmidrule(lr){3-4}
     & & CNN & \cite{liu2020towards}\\ \cmidrule(lr){3-4}
     & & GNN & \cite{li2020custom}\\
\bottomrule

    \end{tabular} 
    \end{center} 
\end{table*}

Analog layout is a hard problem because the parasitics in the layout have a significant impact on circuit performances. This leads to a performance difference between pre-layout and post-layout simulations. Meanwhile, the relation between layout and performance is complex. Traditionally, circuit designers estimate parasitics according to their experience, leading to a long design time and potentials for inaccuracies~\cite{ren2020paragraph}. Therefore, automated analog layout has drawn attention from researchers. Recently, the development of machine learning algorithms promotes research on this problem. All the studies introduced below are summarized in \Cref{tab:layout}.



\citet{xu2019wellgan} use GAN to guide the layout generation. The network learns and mimics designers' behavior from manual layouts. Experiments show that generated wells have comparable post-layout circuit performance with manual designs on the op-amp circuit. \citet{kunal2020gana} train a GCN to partition circuit hierarchy. The network takes circuit netlist as input and outputs circuit hierarchy. With postprocessing, the framework reaches 100\% accuracy in 275 test cases. \citet{zhang2019circuit} introduce a GNN to estimate Electromagnetic (EM) properties of distributed circuits. And they inversely use the model to design circuits with targeted EM properties. \citet{zhu2019genius} propose a fully automated routing framework based on the variational autoencoder (VAE) algorithm. \citet{wu2015analog} design a knowledge-based methodology. They compare the targeted circuit with legacy designs to find the best match. Meanwhile, they expand the legacy database when new circuits are designed. \citet{liu2020closing} put forward a closed-loop design framework. They use a multi-objective Bayesian optimization method to explore circuit layout and use simulation results as the feedback.

To close the gap between pre-layout and post-layout simulations, some researchers attempt to estimate parasitics before layout. \citet{ren2020paragraph} use GNN to predict net parasitic capacity and device parameters based on the circuit schematic. \citet{shook2020mlparest} define several net features and use a random forest to regress net parasitic resistance and capacity. They also model the multi-port net with a star topology to simplify the circuits. Experiments show that with estimated parasitics, the error between pre-layout and post-layout circuit simulation reduces from 37\% to 8\% on average.

Typically, post-layout simulations with SPICE-like simulators are time-consuming. So many researchers focus on layout performance prediction with ML algorithms. \citet{li2020exploring} compare the prediction accuracy of three classical ML algorithms: SVM, random forest, and nerual network.
They also combine the performance prediction algorithms with simulated annealing to fulfill an automated layout framework.
\citet{liu2020towards} propose a 3D CNN for circuit inputs.
First, circuits are converted to 2D images.
Then a third coordinate channel is added to the image to form 3D inputs.
\citet{li2020custom} propose a customized GNN for performance prediction.
They report a higher accuracy than the CNN-based method~\cite{liu2020towards}.

\subsection{Conclusion of Analog Design}

The power of machine learning algorithms has been demonstrated extensively for analog device sizing, topology design and layout problems. Compared to previous optimization-based algorithms, machine learning methods require fewer simulation rounds but achieve higher quality designs. However, existing methods cannot replace human experts yet in the analog design flow. One obstacle is that the models are learned from a limited dataset and have limited flexibility. Most researchers train and test their method on typical circuits like OTAs. A generalizable model designed for a variety of circuits is desired in the future study. Another challenge is that the vast space of system-level design has not been studied. The potential of machine learning in analog design may be further exploited in the future.

\section{Verification and Testing}
\label{sec:test}

\revise{Verification and testing of a circuit are complicated and expensive processes due to the coverage requirements and the high complexity.
Verification is conducted in each stage of the EDA flow to ensure that the designed chip has correct functions.
On the other hand, testing is necessary for a fabricated chip.
Note that from many perspectives, verification and testing share common ideas and strategies, meanwhile face similar challenges.
For instance, with the diversity of applications and the complexity of the design, traditional formal/specification verification and testing may no longer meet various demands.
}

\revise{For the coverage requirements, }
a circuit or system can be very complex and may have many different functions corresponding to different input data. To \revise{verify} a system with low cost, the test set design should be compact and avoid containing ``repeated'' or ``useless'' situations with covering enough combinations of inputs to ensure \revise{reliability}. Therefore, a well-selected test set and a proper strategy are crucial to the fast and correct \revise{verification}.
Traditionally, \revise{for test set design, random generation algorithms and Automated Test Pattern Generation (ATPG) are usually used in the verification stage and the testing stage, respectively.} And their designs \revise{are} always far from the optimal solution. Therefore, it is intuitive to optimize the \revise{verification} process by reducing the redundancy of the test set.

\revise{High complexity of chip testing/verification is another problem. For example, in the analog/RF system design, it is expensive and time-consuming to test the performance accurately or to verify the SPICE netlist formally. Predicting accurate results with low precision test results derived from cheap testing methods is a promising solution to this problem.} 

To meet the coverage requirements and reduce complexity, more and more ML algorithms are applied \revise{in the verification and testing process}, to make fast analog/RF system testing, build simplified estimation model, infer and predict the verification results, optimize sample strategies, and even generate high quality test benches.
These methodologies can be divided into two categories: 1) Machine learning for test set redundancy reduction, \revise{which is applied in both verification and testing stage};
2) Machine learning for complexity reduction, which is applied in \revise{chip testing, verification, and diagnosis}.

\subsection{Machine Learning for Test Set Redundancy Reduction}

{Coverage is the primary concern when designing a test set in \revise{verification and testing} problems. However, the definition of ``coverage'' is different in different problems. For example, for digital design, the test set is supposed to cover as many states of the finite state machine (FSM) or input situations as possible. For analog/RF design, since the input is continuous and the system can be very sensitive to environmental disturbance, a sampling strategy that can cover most input values and working situations is needed. As for the test of semiconductor technology, a test point is a design that needs to be synthesized or fabricated, and the test set needs to cover the whole technology library. We will introduce these problems and corresponding studies based on ML techniques.}

\subsubsection{Test Set Redundancy Reduction for Digital Design Verification}
{\revise{The verification of a digital design will be carried out in each stage of the EDA flow, in which the verification space} of a digital design under test (DUT) usually includes a huge number of situations. Thus, manually designing the test set requires rich expertise and is not scalable. Originally, the test set is usually generated by a biased random test generator with some constraints~\cite{test_survey}, which can be configured by setting a series of directives. Later on, Coverage-Directed test Generation (CDG) techniques have been explored to optimize the test set generation process. The basic idea of CDG is to simulate, monitor, and evaluate the coverage contribution of different combinations of input and initial state. And then, the derived results are used to guide the generation of the test set. There are many CDG works that are based on various ML algorithms such as Bayesian Network~\cite{1219010}, Markov Model~\cite{4167997}, Genetic Algorithm~\cite{4711612,1656859}, rule learning~\cite{7001424,5982005,Eder_ILP}, SVM~\cite{6386595,4555820} and NN~\cite{Wang_GLSVISI18}. We refer the readers to \cite{test_survey} for a more detailed survey on related papers before 2012.} \revise{Note that although the word ``test'' is mentioned frequently in this field, these works mainly aim at aiding the verification process.}

{GA can be applied in CDG problems. \citet{4711612} combine the biased random test generation with GA. First, a constraint model is described and encoded, then a set of constraint models with different configurations is sent into the simulator to \revise{evaluate} the coverage performance. GA method is used to search for a better configuration with higher coverage. \citet{1656859} propose a high-level hardware modeling methodology to get a better description of FSM states and use GA to find a proper configuration of the test generator.}

{Beyond the traditional search strategies, more studies incorporated ML-based models to guide the search process. \citet{6386595} use a one-class SVM for novel test detection. They assume that novel test instances are more useful and could cover more specific corners, and a one-class SVM is used to find these novel instances. \citet{4555820} conduct the functional test selection by using unsupervised Support Vector Analysis. The basic idea is to cluster all the input operations into several groups (e.g., AND operation, other logic operation, and all other operations). Then, one can select the relevant test subsets for specific functional \revise{verification}. A recent study~\cite{Wang_GLSVISI18} focuses on clustering input instructions and adopts an ANN-based method to decide whether a single input instruction should be \revise{verified}.}

{Probabilistic models are also adopted to model the DUT behavior or the test generator. \citet{1219010} propose a CDG method by building a Bayesian Network between the test generator directives and coverage variables. To model the influence of input, some hidden layers are added to the network with expert domain knowledge to help explain the relationship. The Bayesian Network is dynamic and can be adjusted according to stimuli results to get a more precise model. 
Then, we can change the directives to achieve better coverage by running inference on the Bayesian Network. 
Markov model is a special case of Bayesian Network, and \citet{4167997} propose a Markov model for more efficient microprocessor verification. The proposed Markov model shows the transfer probability of different types of instructions. Activity monitors are used for coverage estimation, and the results are used for the adjustment of the Markov model.
The learned Markov model is used as the test generator to achieve better coverage performance.}

{To extract more interpretable and compact knowledge from previous \revise{verification} experiences, rule learning techniques also play a role in CDG problems.
\citet{5982005} apply a decision tree for rule learning of microarchitecture behaviors.
\citet{Eder_ILP} adopt the inductive logic programming method to discover instruction rules, which can be directly used as the directives for further test generation. 
\citet{7001424} propose to discover a subgroup of states that differentiate the failing test cases from the success test cases. All these methods aim at extracting the internal rules of the \revise{verification} problem. With the extracted rules, one can generate better test instances either manually or automatically.}

\subsubsection{Test Set Redundancy Reduction for Analog/RF Design \revise{Testing}}
The Analog/RF system testing can be divided into two aspects, including device-level testing and circuit-level testing. 
The current practice for testing an Analog/RF system is specification testing~\cite{4358314}. This method needs to measure the parameters of the circuit directly. The device will be continuously switched to various test configurations during its operation, resulting in a long setup and establishment time. In each test configuration, measurements are performed multiple times and averaged to reduce thermal noise and crosstalk. Moreover, this complex process needs to be repeated in various modes of operation, such as temperature, voltage level, and output load. Therefore, despite the highly accurate measurement, the overall test process is extremely costly. On the other hand, specification testing requires the use of automatic test equipment (ATE), and the cost of this equipment is also very high. 

A direction to solve these problems is to identify and eliminate the information redundancy in the test set by machine learning, and make a pass/fail decision only depending on a subset of it~\cite{4209884,5169847}. In the specification test, each performance parameter may have redundant information. However, this information needs advanced statistical methods to obtain. ML can help find the complex association in the specification test, so as to reduce the types and times of the test set, and finally complete the result inference with high quality. A multi-objective genetic algorithm is applied for feature selection of the test set, which is used to extract subsets and build the prediction model based on a binary classifier to determine whether the equipment is qualified~\cite{5169847}. The classifier can be constructed by {kNN} or Ontogenic Neural Network (ONN). Taking the power set as an example, the results show that a relatively small number of non-RF specification tests (i.e., digital, DC, and low frequency) can correctly predict a large proportion of pass/fail tags. The experimental results also show that adding some RF specification tests can further improve the prediction error.
 
\begin{figure}[tbp!]

        \centering
        \includegraphics[width=0.6\linewidth]{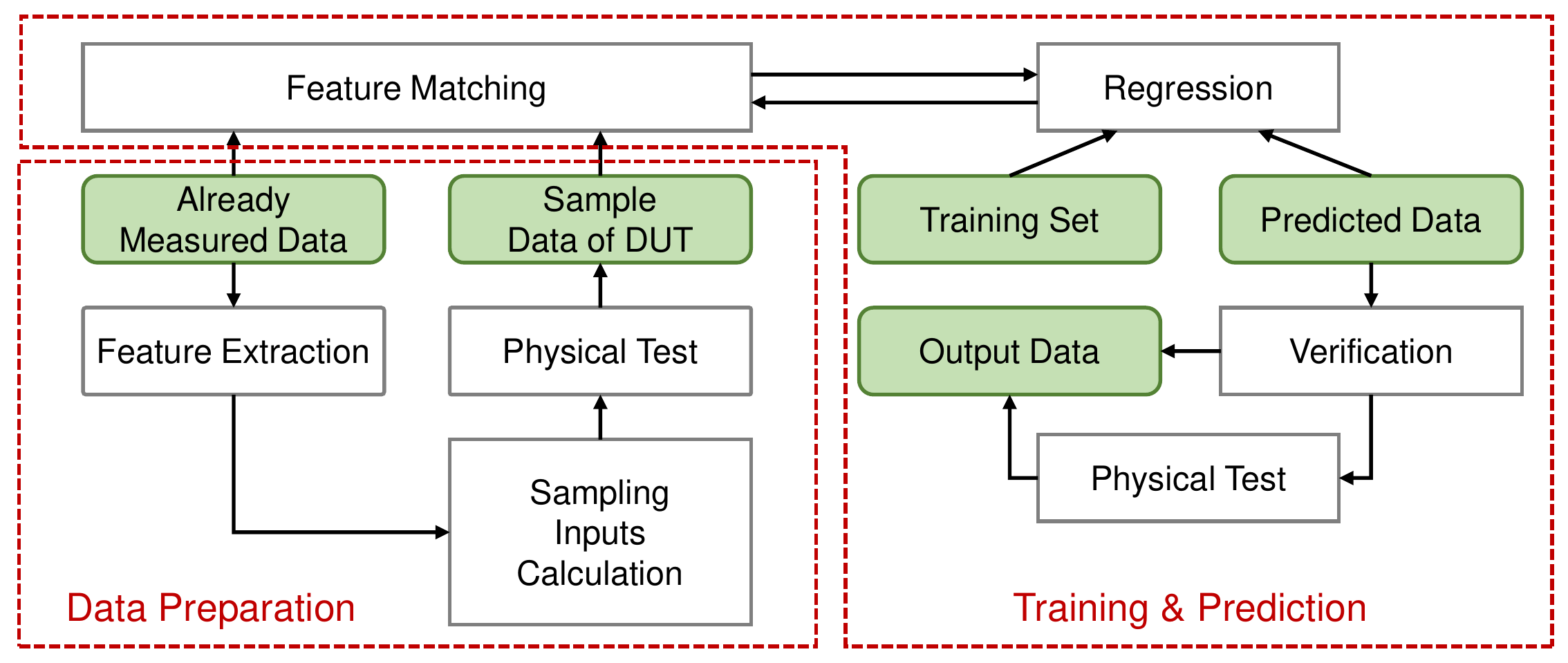}
        \caption{The framework of \cite{8342115} \revise{(reproduced from \cite{8342115})}.}
        \label{test_fig5}

\end{figure}

\citet{8342115} propose a low-cost characterization method for IC technologies. They assume the devices on different dies have similar characteristics, and it is possible to use part of test samples to predict the detailed data. The framework of this work is shown in \Cref{test_fig5}. A small number of samples are tested, and several features are extracted from the test results. Then, the features are used to fit a regression model, with which one can infer the performance curve and predict test results of other samples. In the experiment, the authors use 267 data samples to predict 3241 data points with 0.3\% average error, which reaches a 14x speedup in the test process.

\subsubsection{Test Set Redundancy Reduction for Semiconductor Technology Testing}
\begin{figure*}[htbp]
      \centering
      \includegraphics[width=0.80\linewidth]{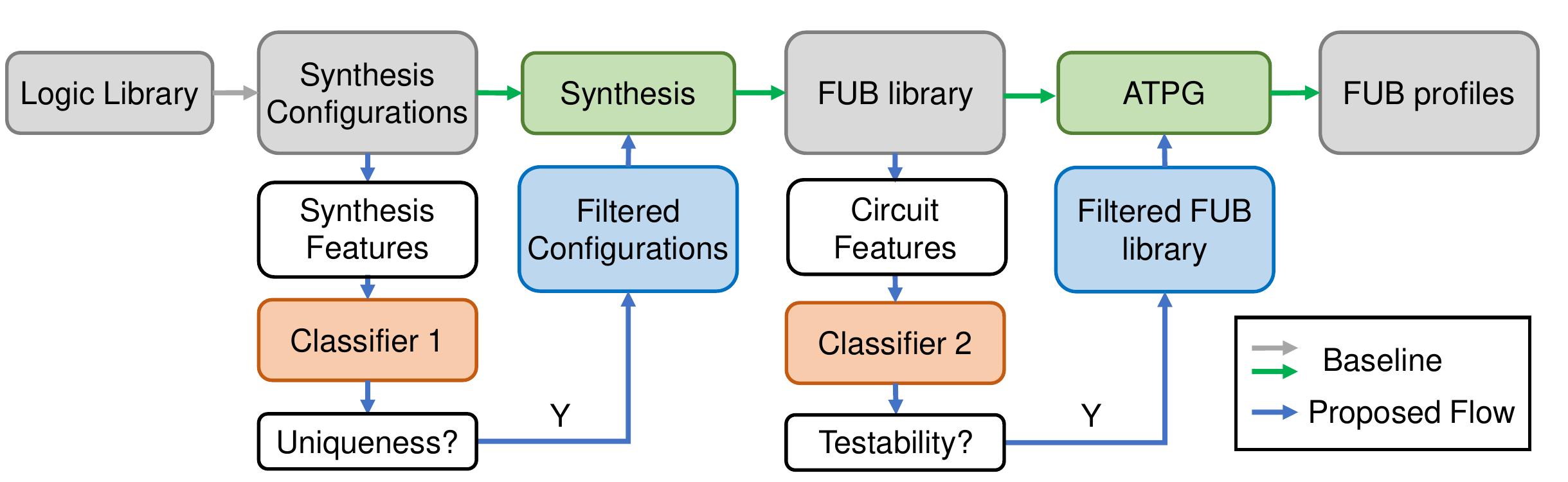}
      \caption{The flow of proposed method in \cite{9000131} \revise{(reproduced from \cite{9000131})}.}
      \label{test_fig4}
  \end{figure*}

Sometimes the problem is to test a new semiconductor technology rather than a specific design.
In this situation, a test instance is a synthesized or fabricated chip design, and building a test set can be extremely expensive.
This problem is a little different from the testing problems mentioned before, but the idea of reducing the test set redundancy is still working.
If we can predict the test set quality and select good parts in advance, the cost can be reduced significantly.
\citet{9000131} focus on optimizing the test set design via ML, of which the proposed flow is shown in \Cref{test_fig4}.
In a traditional testing flow, every possible configuration in a logic library is synthesized, which causes huge time and energy consumption.
To alleviate the problem, this work uses RF models to predict whether a test datum is ``unique'' and ``testable'' with several features
(e.g., the number of nets, fanout, and max logic depth).
``Unique'' means that the test data has a different logical structure compared to other test data, and ``testable'' means that this test data can cover a great number of IP fault.
The experimental results show that this work can achieve over 11$\times$ synthesis time reduction.
  
\subsection{Machine Learning for Test \& Diagnosis Complexity Reduction}

\subsubsection{\revise{Test Complexity Reduction for Digital Design}}
Recently, {GCNs} are used to solve the observation point insertion problem \revise{for the testing stage}~\cite{8807085}. Inserting an observation point between the output of module 1 and the input of module 2 will make the test results of module 1 observable and the test inputs of module 2 controllable. \citet{8807085} propose to use GCN to insert fewer test observation points while maximizing the fault coverage. More specifically, the netlist is first mapped to a directed graph, in which nodes represent modules, and edges represent wires. Then, the nodes are labeled as easy-to-observe or difficult-to-observe, and a GCN classifier is trained. Compared with commercial test tools, this method based on GCN can reduce the observation points by 11\% under similar fault coverage, and reduce the test pattern count by 6\%. Note that compared with other studies discussed before, observation point insertion reduces the test complexity in a different way, by decoupling the test of different modules.

\subsubsection{\revise{Verification} Diagnosis Complexity Reduction for Digital Design}

{During the \revise{verification} process, a complicated diagnosis is needed whenever a bug is detected. However, this diagnosis process might be redundant sometimes since there are lots of similar bugs caused by the same hardware problem, and one situation can be analyzed repeatedly. To alleviate this problem, \citet{7827694} propose an automatic hardware diagnosis method named BugMD, which can classify different bugs and localize their corresponding module. With this framework, the emerging bugs can be analyzed without a complicated diagnosis process. First, the instruction windows containing bugs are encoded to input feature vectors based on the mismatch between DUT and a golden instruction set simulator, then the feature vectors are sent to a classifier for further triaging and localizing, where the ML algorithm can be a decision tree, RF, SVM or NN. To produce sufficient training data, a synthetic bug injection framework is proposed, which is realized by randomly change the functionality of several modules. The experimental results prove the feasibility of BugMD with over 90\% top-3 localization accuracy.} 

\subsubsection{\revise{Verification} \& Test Complexity Reduction for Analog/RF Design}

With increasing system complexity and rising demand for robustness, Analog/RF signal verification has become a key bottleneck~\cite{4167767}, which makes failure detection and design verification very challenging.

A feasible way to reduce the cost of Analog/RF system verification is to use low-cost test equipment to obtain simple results.
Then ML models can be used to map from simple results to complex results obtained by specification testing~\cite{4079383,1443430}.
The basic assumption is that the training set reflects the statistical mechanisms of the manufacturing process,
thus the learned mapping can generalize for new device instances.
Nevertheless, the ML model might fail to capture the correct mapping for some devices since the actual mapping is complex and is not a one-to-one mapping.
Thus, a two-tier test method combining machine learning and specification testing is proposed to improve the accuracy of results~\cite{4358314}. 
During the process, the equipment is first tested by low-cost machine learning-based testing, and the reliability of the results is evaluated. If it is considered insufficient, the more expensive specification testing is conducted. An Ontogenic Neural Network (ONN) is designed to identify the ambiguous regions, and forward the devices to the specification testing. This two-tier approach achieves a trade-off between accuracy and cost.

\begin{figure}[htbp]
\centerline{\includegraphics[height=4cm,width=8cm]{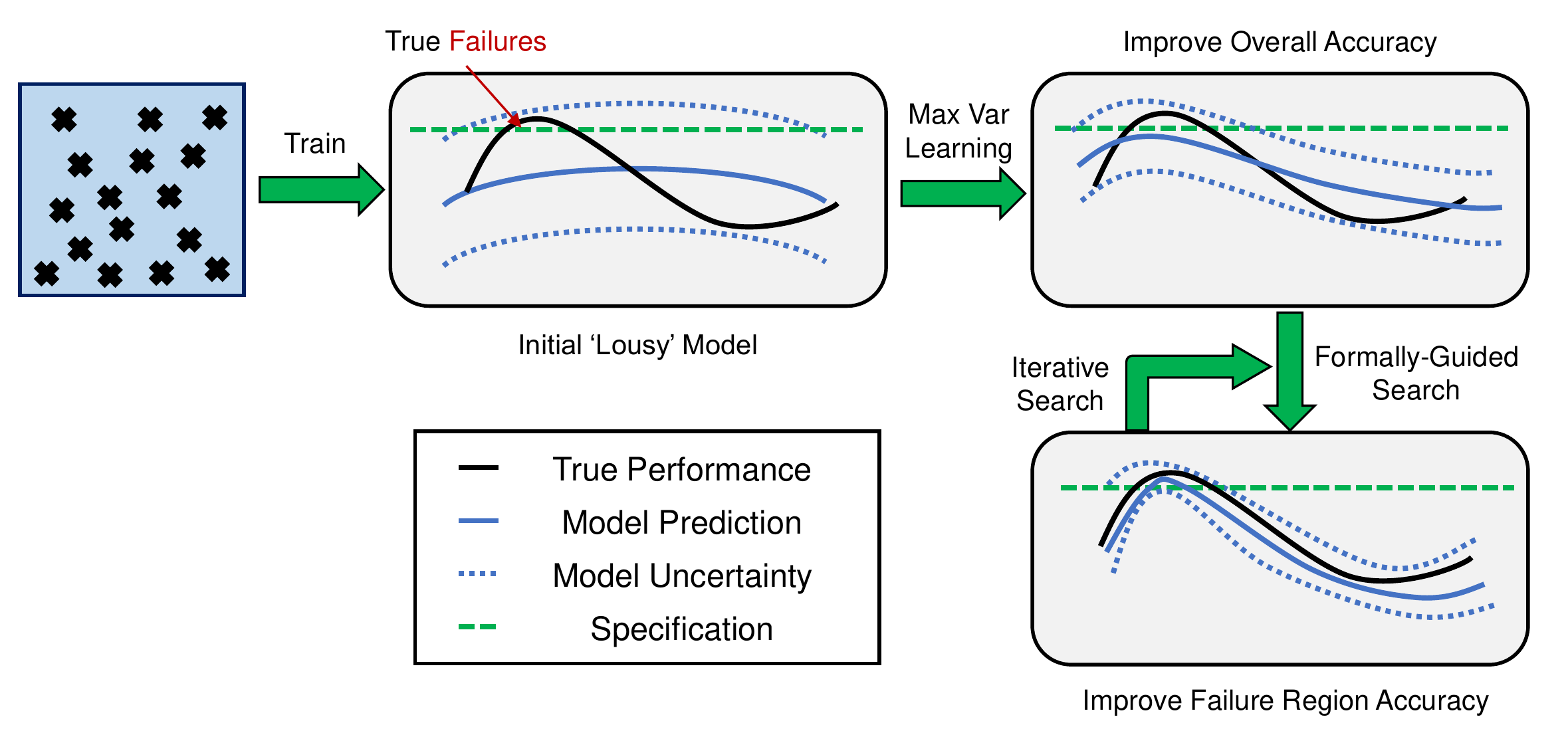}}
\caption{Active learning for circuits testing (reproduced from~\cite{8465826}).}
\label{test_fig2}
\end{figure}

\revise{Although formal verifications can provide guarantees for the specifications under check, they are only feasible for small analog blocks with idealistic models and fail for practical usage on large detailed SPICE circuit netlist. Therefore, machine learning is applied to aid the verification process.} HFMV~\cite{8465826} combines a machine learning model with formal verification: When there is insufficient confidence in the test results of the machine learning model, formal verification is performed. HFMV proposes a probabilistic machine learning model to check whether there is enough confidence to meet the target specification.
As shown in \Cref{test_fig2}, 
HFMV relies on two active learning approaches to improve the performance of the ML model, including 1) max variance learning to reduce model uncertainty; 2) formally-guided active learning to discover rare failure regions. Their results show that HFMV can detect \revise{rare failures}.

\subsection{Summary of ML for Verification and Testing}
{There are mainly two ways of accelerating the \revise{verification and testing} process: \revise{1) Reducing the test set redundancy; 2) Reducing the complexity of the testing, verification and diagnosis process}. To reduce the test set redundancy or to optimize the generation of test instances, coverage-directed test generation has been studied for a long time, which can be aided by lots of ML algorithms. Recently, test set redundancy reduction of analog/RF design or even the test of semiconductor technology have raised a lot of attention, and more ML methods are applied to solve these problems. As for reducing the \revise{verification \& test complexity}, there are studies that adopt low-cost tests for analog/RF design, and some other studies that focus on fast bug classification and localization. The related works on ML for \revise{verification} \& testing problems are summarized in \Cref{test_summary}.}

\begin{table*} 
    \caption{{Summary of ML for verification and testing} }
    \label{test_summary}
    \begin{center} 
    	\footnotesize
\begin{tabular}{p{3cm}|p{3cm}|p{3cm}|p{3cm}} 
\toprule
     \textbf{\centering{Optimization Idea}} & \textbf{Task} & \textbf{ML Algorithm} & \textbf{References}\\ \midrule
     \multirow{13}{3cm}{\centering Test Set Redundancy Reduction} 
     & \multirow{7}{3cm}{\centering Digital Design} & Statistical Model & \cite{1219010}, \cite{4167997} \\ \cmidrule(lr){3-4} 
     &  & Search Methods & \cite{4711612}, \cite{1656859}\\ \cmidrule(lr){3-4}
     &  & Rule Learning & \cite{7001424}, \cite{5982005}, \cite{Eder_ILP} \\ \cmidrule(lr){3-4}
     &  & CNN, SVM, et al. & \cite{6386595}, \cite{4555820}, \cite{Wang_GLSVISI18}\\ \cmidrule(lr){3-4}
     &  & GCN & \cite{8807085}\\ \cmidrule(lr){2-4}
     & \multirow{2}{3cm}{\centering Analog/RF Design} & KNN, ONN & \cite{5169847}\\ \cmidrule(lr){3-4} 
     &  & Regression & \cite{8342115}\\ \cmidrule(lr){2-4}
     & \centering{Semiconductor Technology} & CNN & \cite{9000131}\\ \cmidrule(lr){1-4}
     \multirow{4}{3cm}{\centering Test Complexity Reduction} & \centering Digital Design & SVM, MLP, CNN, et al. & \cite{7827694}\\ \cmidrule(lr){2-4} 
     & \multirow{2}{3cm}{\centering Analog/RF Design} & ONN & \cite{4358314}\\ \cmidrule(lr){3-4} 
     &  & Active Learning & \cite{8465826}\\  
\bottomrule

    \end{tabular} 
    \end{center} 
\end{table*}

\section{Other Related Studies}
\label{sec:misc}
\subsection{Power Prediction} \label{sec:misc:power}

Power estimation is necessary in electronic system design, which can be carried out at different levels according to application scenarios.
In general, there is a tradeoff between the power estimation accuracy and simulation method complexity.
For example, the gate-level estimation can generate a cycle-by-cycle power track with high accuracy but has a huge time consumption.
In contrast, high-level simulation can only provide less accurate evaluation, but requires less specific knowledge and computing complexity at the same time.
Nevertheless, it is possible for ML methods to make accurate and detailed power prediction only with high-level evaluation,
which shows significant benefits for fast chip design and verification.

\citet{Todaes2018} propose a multi-level power modeling method, which only uses high-level C/C++ behavior description and some hardware information to obtain power model at different granularities.
The derived power model granularity depends on how much information we have about the hardware design, i.e., black, grey, or white box modeling.
For each modeling problem, an evaluation flow is designed, and several regression algorithms are applied.
The proposed flow achieves a significant speedup compared with traditional RTL-level or gate-level simulation within 10\% error.

\citet{Simmani} propose an RTL-level power prediction framework named SIMMANI with signal clustering and power model regression. All the signals are encoded according to the toggle patterns observed in a specific window, which are then clustered and selected. The regression model takes the selected signals as the input, and outputs the power estimation result.

{Besides traditional regression methods, other ML methods also show great potential in power predicting problems. PRIMAL~\cite{Primal} is an RTL power estimation framework based on several ML methods, including {Principal Component Analysis (PCA)}, MLP and CNN. In PRIMAL, the toggle patterns of registers are first encoded into 1D or 2D features and then processed by various ML algorithms. The trained model can evaluate the power track for new workloads that are very different from the training set. To enhance the local information, a graph-based partitioning method is leveraged for the mapping strategy from registers to feature pixels. PRIMAL can achieve a 50$\times$ speedup than gate-level power estimation flow with an average error below 5\%.} 

{PRIMAL is a promising solution to RTL power prediction. However, there exist transferability problems with this solution in that a power model can only describe a specific system design. That is to say, we have to train a new model for a new design. To solve this problem, a GNN-based framework named GRANNITE is proposed by \citet{Grannite}. Different from PRIMAL, GRANNITE takes the gate-level netlist into consideration to build a GNN. GRANNITE shows good transferability among different designs by utilizing more hardware details. Note that this work still conducts an RTL-level power prediction, since the gate-level netlist is only used for the graph generation, and no gate-level power estimation is involved. Compared to a traditional probabilistic switching activity estimation, GRANNITE achieves a speed up of two orders of magnitude on average, and the average relative error is within 5.5\%.}

\subsection{Machine Learning for SAT Solver}
SAT plays an important role in circuit design and verification, error diagnosis, model detection of finite state machines, FPGA routing, logic synthesis and mapping, register allocation, timing, etc. Researchers contribute to improving the efficiency of the search engine in SAT solvers and design various strategies and heuristics. Recently, with the advancement of NNs in representation learning and solving optimization problems, there have been increasing interests in generating and solving SAT formula with NNs. 


The performance of the conflict-driven Davis Putnam style SAT solver largely depends on the quality of restart strategies. \citet{b41} successfully apply a supervised learning method to design LMPick, a restart strategy selector. Among various heuristics, branching heuristics~\cite{b17,b18,b19,b20} attract lots of attention for its great performance. Multi-class SVM is applied in \cite{b21} to tune parameters of heuristics, according to the features of both input and output clauses. SATzilla~\cite{b22} integrates several solvers and builds an empirical hardness model for solver selection.
Some work~\cite{b33,b34,b35} evolve heuristics through genetic algorithms by combining existing primitives, with the latter two aiming at specializing the created heuristics to particular problem classes. There have also been other approaches utilizing reinforcement learning to discover variable selection heuristics~\cite{b36,b18,b38,b39,b40}.

Recently, NNs have found their applications in solving SAT.
\citet{b23} introduce the recurrent relational network to solve relational inference, e.g.~Sudoku.
\citet{b24} present an NN architecture that can learn to predict whether one propositional formula entails another by randomly sampling and evaluating candidate assignments. 
There have also been several recent papers showing that various neural network architectures can learn good heuristics for NP-hard combinatorial optimization problems~\cite{b25,b26,b27}.
\citet{b28} propose to train a GNN (called NeuroSAT) to classify SAT problems as satisfiable or unsatisfiable.
\citet{b29} also use a simplified NeuroSAT to guide the search process of an existing solver.

In recent studies, a common practice is to use GNN for feature extraction and reinforcement learning for learning the policy.
\citet{b30} learn improved heuristics to solve quantified Boolean formulas via reinforcement learning while using GNN for formula encoding. \citet{yolcu2019learning} also use RL to learn local search heuristics with a GNN serving as the policy network for variable selection. Besides GNN, RNN can also be employed for formula or DAG embedding.
Lately, \citet{b31} propose Circuit-SAT to solve SAT problems, employing gated recurrent units that can implement sequential propagation of DAG-structured data.
The training procedure works in the exploration and exploitation manner, which is similar to the reinforcement learning paradigm.

\subsection{Acceleration with Deep Learning Engine}
EDA tools typically involve solving large-scale optimization problems with heavy numerical computation, especially at the physical design stage, and extensive work is devoted to accelerating these solvers with modern parallel computing hardware like multicore CPUs or GPUs~\cite{cheng2018replace,lu2015eplace,cong2009parallel}. {Many recent studies have explored GPU's opportunity in EDA problems~\cite{lin2020dreamplace,10.1145/3400302.3415750,10.1145/3400302.3415773,10.1145/3400302.3415762}. Still, developing good GPU implementation of EDA algorithms is challenging.}

\citet{lin2020dreamplace} leverage the mature deep learning engines to build a GPU-accelerated placement framework called DREAMPlace Advancement in ML has encouraged the development of software frameworks and tool-kits which decouple algorithmic description from system implementation (e.g., interaction with GPUs, optimizing low-level operator code) to help develop ML models productively ~\cite{abadi2016tensorflow,paszke2017automatic}. The key insight of this paper is that the analytical placement problem is analogous to the training of a NN model. They both involve optimizing some parameters (i.e., cell locations in placement, weights in NN) to minimize a cost function (i.e., wirelength in placement, cross-entropy loss in NN). With hand-optimized key operators integrated in DL training framework PyTorch, DREAMPlace demonstrates over $40\times$ speedup against CPU-based multi-threaded tools~\cite{lu2015eplace,cheng2018replace}. The tool claims to be extensible to new solvers by simply adding algorithmic description in high-level languages like Python. 

\subsection{Auto-tuning design flow}
With the increasing complexity of chip design, massive choices and parameters of the synthesis tools make up huge design space.
To improve the efficiency of tuning, recent studies employ more advanced learning-based algorithms.
In \cite{Auto2016}, some complete parameter settings are selected and then gradually adapted during synthesis to achieve optimal results.
\citet{Auto2019} propose the first recommender system based on the collaborative filtering algorithm.
The system consists of two modules: the offline learning module and the online recommendation module.
The offline learning module is to predict QoR given macro specification, parameter configuration, cost function and iterative synthesis output.
The online recommendation module generates several optimal settings.
A recent study~\cite{FIST} also employs a tree-based XGBoost model for efficient tuning.
Besides, this paper also designs a clustering technique that leverages prior knowledge and an approximate sampling strategy to balance exploration and exploitation.
In \cite{RLplacementParamTuning}, a deep RL framework that adopts unsupervised GNN to generate features is developed to automatically tune the placement tool parameters.

\graphicspath{{image/ml/}}

\section{Discussion From the Machine Learning Perspective}
\label{sec:ml}
In this section, we revisit some aforementioned research studies from an ML-application perspective.

\subsection{{The Functionality of ML}} \label{sec:ml:models}

\begin{table*} 
    \caption{{Overview of ML functionality in EDA tasks}}
    \label{ml-model}
    \begin{center} 
        \scriptsize
        \centering
\begin{tabular}{p{2cm}|p{2.5cm}|p{2cm}|p{2cm}|p{2cm}|p{1.5cm}}
\toprule
    \textbf{ML Functionality} & \textbf{Task / Design Stage} & \textbf{ML Algorithm} & \textbf{Input} & \textbf{Output} & \textbf{Section}\\ \midrule
    
    \multirow{4}{2cm}{Decision making in traditional methods}
    
    & HLS Design space exploration & Decision Tree, quadratic regression, etc. & Hardware directives (pragmas) in HLS design & Quality of hyper-parameters, e.g., initial state, termination conditions & 
    \Cref{sec:hls:dse:traditional} 
    \\ \cmidrule(lr){2-6} 
    
    & Logic synthesis & DNN & RTL descriptions & Choice of the workflow and optimizer  & \Cref{sec:physical:logicsynthesis}
    \\ \cmidrule(lr){2-6} 
    
    & Mask synthesis & CNN &  Layout images & Choice of optimization methods & \cite{mask-optimization} in \Cref{sec:mask:opc} \\ 
    \midrule

    & Analog topology design & CNN, Fuzzy logic, etc. & Analog specifications & Best topology selection & \Cref{sec:analog:topologyselection}\\ \cmidrule(lr){2-6} 

    \multirow{5}{2cm}{\centering Performance prediction} 
    & HLS & Linear Regression, SVM, Random Forest, XGBoost, etc. & HLS Report, workload characteristics, hardware characteristics & Resource usage, timing, etc. & \Cref{sec:hls:prediction}
    \\ \cmidrule(lr){2-6} 
    
    & Placement and routing & SVM, CNN, GAN, MARS, Random Forest etc. & Features from netlist or layout image & Wire-length, routing congestion, etc. & \Cref{sec:physical:PnR} \\ \cmidrule(lr){2-6} 

    & Physical implementation (lithography hotspot detection, IR drop prediction, power estimation, etc.) & SVM, CNN, XGBoost, GAN, etc. & RTL and gate-level descriptions, technology libraries, physical implementation configurations & Existence of lithography hotspots, IR drop, path delay variation, etc & \Cref{sec:lithography_hotspot}--\ref{sec:physical:other}, \ref{sec:mask:lithosimulation}, \ref{sec:misc:power}\\ \cmidrule(lr){2-6} 

    & Verification & KNN, Ontogenic Neural Network (ONN), GCN, rule learning, SVM, CNN & Subset of test specifications or low-cost specifications & boolean pass/fail prediction & \Cref{sec:test}\\ \cmidrule(lr){2-6}
    
    & Device sizing & ANN & Device parameter & Possibility of constraint satisfaction & \Cref{sec:analog:devicesizing}\\ 
    
    \toprule 
    
    \textbf{ML Functionality} & \textbf{Task / Design Stage} & \textbf{ML Algorithm} & \textbf{Tuning parameters} & \textbf{Optimization Objective} & \textbf{References}\\ \midrule
    \multirow{2}{2cm}{\centering Black-box optimization} 
    & HLS Design Space Exploration & Random Forest, Gaussian Process, Ensemble models, etc. & Hardware directives (pragmas) in HLS design & Quality-of-Results, including latency, area, etc. & \Cref{sec:hls:dse:learning}
    \\ \cmidrule(lr){2-6} 

    & 3D Integration & Gaussian Process, Neural Network & Physical design configurations & Clock skew, thermal performance, etc. & \Cref{sec:physical:3d}
    \\ \midrule

    \multirow{4}{2cm}{\centering Automated design} 
    & Logic synthesis & RL, GCN & Gate-level DAG for a logic function & Area, latency, etc. & \Cref{sec:physical:logicsynthesis}
    \\ \cmidrule(lr){2-6}

    & Placement & RL, GCN & Macro placement position & Wire-length, congestion, etc. & \cite{mirhoseini2020chip} in \Cref{sec:physical:PnR:placementdecision}\\ \cmidrule(lr){2-6} 

    & Mask synthesis & GAN, CNN, Decision Tree, dictionary learning, etc. & RTL and gate-level description, layout images & Generated optical proximity correction (OPC) and sub-resolution assist feature (SRAF) & \Cref{sec:mask:opc}--\ref{sec:mask:sraf} \\\cmidrule(lr){2-6} 

    & Device sizing & RL, GCN, DNN, SVM & Device parameters & Satisfaction of design constraints & \Cref{sec:analog:devicesizing}
    \\ \bottomrule
    \end{tabular} 
    \end{center} 
\end{table*}

\Cref{bg:ml} introduces the major ML models and algorithms used in EDA problems.
{Based on the functionality of ML in the EDA workflow,} we can group most researches into {four} categories:
decision making in traditional methods, performance prediction, black-box optimization, and automated design. 

{\underline{\textit{Decision making in traditional methods.}} The configurations of EDA tools, including the choice of algorithm or hyper-parameters, have a strong impact on the efficiency of the procedure and quality of the outcome.
This class of researches utilizes ML models to replace brute-force or empirical methods when deciding configurations. ML has been used to select among available tool-chains for logic synthesis~\cite{LSOracle,CNNlogic} , mask synthesis~\cite{mask-optimization}, and topology selection in analog design~\cite{fasy_tcad1996,fuzzy_,inference_sisa2018}. ML has also been exploited to select hyper-parameters for non-ML algorithms such as Simulated Annealing, Genetic Algorithm, etc. (refer to \Cref{sec:hls:dse:traditional}).}

\underline{\textit{Performance prediction.}} This type of tasks mainly use supervised or unsupervised learning algorithms. 
Classification, regression and generative models are trained by former cases in real production to estimate QoR rapidly,
to assist engineers to drop unqualified designs without time-consuming simulation or synthesis. 

ML-based performance prediction is a very common type of ML application. 
Typical applications of this type include congestion prediction in placement \& routing and hotspot detection in manufacturability estimation (\Cref{ml-model}). 
The most commonly-used models are Linear Regression, Random Forests, XGBoost, and prevailing CNNs. 

\underline{\textit{Black-box optimization.}} This type of tasks mainly use active learning.
Many tasks in EDA are DSE, i.e., searching for an optimal (single- or multi-objective) design point in a design space.
Leveraging ML in these problems usually yields black-box optimization, which means that the search for optimum is guided by a surrogate ML model,
not an explicit analytical model or hill-climbing techniques.
The ML model learns from previously-explored design points and guides the search direction by making predictions on new design points.
Different from the first category, the ML model is trained in an active-learning process rather than on a static dataset,
and the inputs are usually a set of configurable parameters rather than results from other design stages. 

Black-box optimization is widely used for DSE in many EDA problems.
Related ML theories and how to combine with the EDA domain knowledge are extensively studied in literature.
Typical applications of this type include tuning HLS-level parameters and physical parameters of 3D integration (see \Cref{ml-model}).
The key techniques are to find an underlying surrogate model and a search strategy to sample new design points.
Options of the surrogate model include GP, along with all the models used in performance prediction~\cite{ma2018cross,Meng2016}.
Search strategies are usually heuristics from domain knowledge, including uniformly random exploration~\cite{liu2013learning},
exploring the most uncertain designs~\cite{Zuluaga2013}, exploring and eliminating the worst designs~\cite{Meng2016}, etc.

\underline{\textit{Automated design.}} Some studies leverage AI to automate design tasks that rely heavily on human efforts. Typical applications are placement~\cite{mirhoseini2020chip} and analog device sizing~\cite{learning,gcn-rl,autockt}. At first look it is similar to black-box optimization, but we highlight the differences as:
\begin{itemize}
    \item The design space can be larger and more complex, for example in placement, the locations of all the cells. 
    \item Instead of searching in the decision space, there exists \textbf{a trainable decision-making policy} that outputs the decisions,
        which is usually learned with RL techniques.
\end{itemize} 
More complicated algorithms with large volumes of parameters, such as deep reinforcement learning, are used in these problems. This stream of researches show the potential to fully automate IC design.

\Cref{ml-model} summarizes representative work of each category and typical model settings in terms of algorithm, input and output.

\subsection{Data Preparation}

The volume and quality of the dataset are essential to model performance. Almost all studies we review make some discussions on leveraging EDA domain knowledge to engineer a large, fair and clean dataset. 

\underline{\textit{Raw data collection.}} Raw features and ground truth / labels are two types of data needed by ML models. 
Raw feature extraction is often a problem-specific design, but there are some shared heuristics. Some studies treat the layout as images and leverage image processing algorithms~\cite{AENEID,J-net,Routenet}. 
Some choose geometric or graph-based features from the netlist~\cite{pade_placer}. Some use traditional algorithms to generate features~\cite{Routenet,9045178,6681682,8533535}. 
Quite a lot studies choose features manually~\cite{7835438,8394712,9045178,8060438,10.1145/3287624.3287689,7428008,LSOracle}.
To some extend, manual feature selection lacks a theoretical guarantee or practical guidance for other problems. 
The labels or ground truth are acquired through time-consuming simulation or synthesis. This also drives researchers to improve data efficiency by carefully architect their models and preprocess input features, or use semi-supervised techniques~\cite{8695779} to expand the dataset.

\underline{\textit{Feature preprocessing.}}
Standard practices like feature normalization and edge data removal are commonly used in the preprocessing stage.
Some studies also use dimension reduction techniques like {PCA and LDA} to further adjust input features~\cite{8715016}. 

\subsection{Domain Transfer}

There have been consistent efforts to make ML-based solutions more adaptive to domain shift, so as to save training from scratch for every new task. Some researches propose ML models that take specifications of the new application domain and predict results in new domain based on results acquired in original domain. This idea is used in cross-platform performance estimation of FPGA design instances~\cite{moham2019,8587690}. It would be more exciting to train AI agents to adapt to new task without preliminary information of the new domain, and recent studies show that Reinforcement Learning (RL) might be a promising approach. RL models pre-trained on one task is able to perform nicely on new tasks after a fine-tune training on the new domain~\cite{autockt,gcn-rl,mirhoseini2020chip}, which costs much less time than training from scratch and sometimes lead to even better results.

\section{Conclusion and Future Work}
\label{sec:conclu}

It is promising to apply machine learning techniques in accelerating EDA tasks.
In this way, the EDA tools can learn from previous experiences and solve the problem at hand more efficiently.
{So far machine learning techniques have found their applications in almost all stages of the EDA hierarchy.
In this paper, we have provided a comprehensive review of the literature from both the EDA and the ML perspectives.}

Although remarkable progress has been made in the field, we are looking forward to more studies on applying ML for EDA tasks from the following aspects. 
\begin{itemize}

    \item \underline{\textit{Towards full-fledged ML-powered EDA tools}}.
        In many tasks (e.g., analog/RF testing, physical design), the performance of purely using machine learning models is still difficult to meet the industrial needs.
        Therefore, smart combination of machine learning and the traditional method is of great importance.
        Current machine learning aided EDA methods may be still restricted to less flexible design spaces, or aim at solving a simplified problem.
        New models and algorithms are desired to be developed to make the ML models more useful in real applications.

    \item \underline{\textit{Application of new ML techniques}}.
        Very recently, some new machine learning models and methodologies (e.g., point cloud and GCN) and machine learning techniques (e.g., domain adaptation and reinforcement learning)
        begin to find their application in the EDA field.
        We expect to see a broader application of these techniques in the near future.

    \item \underline{\textit{Trusted Machine Learning}}.
        While ML holds the promise of delivering valuable insights and knowledge into the EDA flow, broad adoption of ML will rely heavily on the ability to trust their predictions/outputs.
        For instance, our trust in technology is based on our understanding of how it works and our assessment of its safety and reliability.
        To trust a decision made by an algorithm or a machine learning model, circuit designers or EDA tool users need to know that it is reliable and fair, and that it will cause no harm.
        We expect to see more research along this line making our automatic tool trusted.

\end{itemize}

\section*{Acknowledgment}

This work was partly supported by National Natural Science Foundation of China (No. U19B2019, 61832007, 61621091), and the Research Grants Council of Hong Kong SAR (No.~CUHK14209420).

\bibliographystyle{ACM-Reference-Format}
\bibliography{ref/Top,ref/acmart}

\ifdefined\REVIEW
\clearpage
\pagestyle{empty}
\input{texts/response-v2}
\fi

\end{document}